\documentclass[12pt]{iopart}
\usepackage{iopams}  
\expandafter\let\csname equation*\endcsname=\relax
\expandafter\let\csname endequation*\endcsname=\relaxD
\usepackage{epsf,graphicx,graphics}
\usepackage{amsmath}
\usepackage{latexsym,amssymb}
\usepackage{mathtools}
\usepackage{setspace,cite}
\usepackage{a4}
\usepackage{slashed}
\usepackage{float}
\usepackage[left=3cm,right=3cm,top=3cm,bottom=4cm]{geometry}

\newcommand{\od}[2]{\frac{d #1}{d #2}}

\usepackage{panmaths}

\begin{document}

\renewcommand{\thefootnote}{\arabic{footnote}}
\title[Dyonic furry black holes with $\Lambda<0$]{Dyonic black holes in $\mk{su}(\infty)$ anti-de Sitter Einstein-Yang-Mills theory, characterised by an infinite set of global charges}
\emph{Class. Quantum Grav.} \textbf{36}:18, 185014, 2019. 
\author{J. Erik Baxter}
\address{Physics Research Group,
	Sheffield Hallam University,
	1 Howard Street,
	Sheffield, 
	South Yorkshire, UK
	S11WB}
\ead{e.baxter@shu.ac.uk}
%
%
%
%
\begin{abstract}
	We here derive field equations for static, spherically symmetric, dyonic $\mk{su}(\infty)$ Einstein-Yang-Mills theory with a negative cosmological constant $\Lambda$. We are able to find new non-trivial black hole solutions to this system in two regimes: where the gauge fields are small; and where $|\Lambda|\rar\infty$. We also show that some may be uniquely characterised by a countably infinite set of asymptotically-defined charges. This may have implications for Bizon's modified `No Hair' conjecture.
\end{abstract}
%
\vspace{2pc}

\maketitle

\section{Introduction}\label{Intro}

In the last three decades, `hairy' black holes have been the subject of much interest and many publications. Before this, the uniqueness theorems of Israel, Carter et al. suggested that black holes were very simple objects, characterised entirely by their mass $m$ and their electric charge $e$ (and possibly the angular momentum, for non-static solutions) \cite{israel_300_1987, israel_event_1968, israel_event_1967}. Then, Bizon \cite{bizon_gravitating_1994} and Bartnik and McKinnon \cite{bartnik_particle-like_1988} discovered an infinite family of static black holes and solitons in Einstein-Yang-Mills (EYM) theory, where the symmetry group of the single gauge field is $\essu$. These solutions contrasted with earlier purely gravitational black holes by requiring an extra gauge parameter to fully classify them, as well as their mass and electric charge. This prompted Bizon to propose the following modified ``No Hair'' conjecture:

\begin{quotation}\noindent
	``Within a given matter model, a stable black hole is uniquely characterised by a finite number of global charges.''
\end{quotation}

These asymptotically flat $\essu$ solutions were found to be unstable \cite{volkov_number_1995, zhou_instability_1990, lavrelashvili_remark_1995}, as were solutions in the case of a general gauge group \cite{brodbeck_instability_1996}, thus preserving the spirit (if not the letter) of the original uniqueness theorems. However, solutions for $\essu$ were later found by introducing a negative cosmological constant $\Lambda$, and in addition, these are stable under linear perturbations of the field variables in the limit of large $|\Lambda|$ \cite{winstanley_existence_1999, sarbach_linear_2001, winstanley_linear_2002}. Since then, numerous results have been found for asymptotically Anti-de Sitter (AdS) EYM theories and extensions thereof, including for example non-spherically symmetric models \cite{radu_static_2004, baxter_topological_2016} and models with extra matter fields \cite{winstanley_instability_2016,ponglertsakul_solitons_2016}. Recent works include \cite{baxter_stability_2016,baxter_abundant_2008,baxter_existence_2016,baxter_existence_2018,nolan_existence_2012}; see also \cite{winstanley_menagerie_2015} for a recent review. Furthering the range of available models and testing the modified ``No hair'' theorem is of current interest and is part of the motivation for this work.

One extension of the model that has been of interest recently is to allow the gauge field to possess a non-trivial electric sector -- such solutions include \emph{dyons} and \emph{dyonic black holes}, the analogues of the soliton and black hole solutions we find to purely magnetic field equations. Previous research proved the existence of dyonic solutions for $\sun$, and then for all compact, semi-simple and simply connected gauge groups \cite{baxter_existence_2016, baxter_existence_2018}; and some solutions at least have been shown to be stable \cite{nolan_stability_2016}. This is of great interest since via the AdS/CFT correspondence \cite{maldacena_large_1998}, observables in the CFT are connected to the presence of hair in the dual gravity theory. The correspondence has already been used in the case of certain non-spherically symmetric dyonic models to produce results concerning holographic superconductors \cite{hendi_holographical_2018,cai_introduction_2015,shepherd_black_2017}.

Thus, string theory is also a motivation for this work. It is known that string theories are associated with infinite dimensional Lie algebras, since these correspond to the possible string states. An early infinite dimensional algebra to be investigated was the Virasoro algebra, later extended to the algebra $w_\infty$. It can be shown that $w_\infty$ is isomorphic to the area-preserving diffeomorphisms of $\mbox{SDiff}(\Sigma^2)$ of a string 2D worldsheet $\Sigma^2$; and in the case where $\Sigma^2=S^2$, the 2-sphere, it also can be shown that both of these algebras are also isomorphic to $\suinf$, the limit of $\sun$ as $N\rar\infty$ \cite{bakas_large_1989,sezgin_area-preserving_1992,pope_higher-spin_1990}. Other related motivations for considering the limit $N\rar\infty$ are that the AdS/CFT duality becomes `exact' in this limit in the sense that the string theory will approximate to supergravity on the D-brane as $N\rar\infty$. 
Finally, the question of what such `furry' black holes can teach us about the Black Hole Information Paradox is a subject of current interest \cite{ellis_w_infty_2016,antoniadis_proceedings_2015,hawking_soft_2016,haco_black_2018}.
%

The existence of black hole solutions to the purely magnetic $\suinf$ system has been proven \cite{mavromatos_infinitely_2000}, in the regimes of a small gauge field and a large (negative) value of $\Lambda$; and so that analysis did not include the electric sector. In addition, at that time the broader analytical and numerical behaviour of $\sun$ solutions had not been explored in much detail as it has now \cite{baxter_stability_2016,baxter_existence_2016,baxter_existence_2018}, and hence the question of characterising charges for those solutions (as mentioned in the modified `No Hair' Theorem) was not considered at the time. Inspired by all of the above, we here investigate dyonic black hole and dyon solutions to $\suinf$ field equations, and in doing so revisit the purely magnetic solutions in light of recent research.

\section{Deriving the field equations for $\mk{su}(\infty)$ Einstein-Yang-Mills theory}\label{AnsFE}

In this Section, we set the scene by outlining the derivation of dyonic $\suinf$ field equations. This analysis roughly follows that for purely magnetic field equations as described in \cite{mavromatos_infinitely_2000}, though the difference for us is that we do \emph{not} take a trivial electric sector, which complicates the equations substantially.

We start with the following well-established $\sun$-invariant\footnote{Throughout this work, when we refer to $\sun$, it is to be understood that we are referring to the case where $N$ is finite, as opposed to $\suinf$.} gauge potential \cite{baxter_existence_2016}
\begin{equation}\label{gaugepot}
\begin{split}
\mc{W}\equiv\mc{W}_\mu dx^\mu\equiv& A^{(N)}dt+B^{(N)}dr+\frac{1}{2}\left(C^{(N)}-C^{(N)\dagger}\right)d\theta\\
&-\frac{i}{2}\left[\left(C^{(N)}+C^{(N)\dagger}\right)\sin\theta+D^{(N)}\cos\theta\right]d\phi.
\end{split}
\end{equation}
where $A^{(N)}$, $B^{(N)}$, $C^{(N)}$ and $D^{(N)}$ are matrices in $\sun$ -- $A^{(N)}$ and $B^{(N)}$ are diagonal and imaginary, $D^{(N)}$ is diagonal and real, $C^{(N)}$ is anti-Hermitian and upper-triangular, and all matrices are functions of $r$ alone.

In order to take the limit $N\rar\infty$ sensibly, one must construct specific set of $N^2-1$ generators for $\sun$. The ground for this was laid in \cite{hoppe_quantum_1982}, where it was shown that the structure constants for Poisson-type commutator brackets over a basis of spherical harmonics are identical to those for the commutation relations of $\sun$ in a certain basis of matrix polynomials, and hence that the group $\sun$ in the limit $N\rar\infty$ is isomorphic to the algebra of area-preserving diffeomorphisms on the sphere, $\mbox{SDiff}(S^2)$. This was extended by \cite{floratos_note_1989}, which derived the precise process and dictionary of correspondences that allow one to sensibly consider $\sun$ in the limit $N\rar\infty$, which was in turn applied to the case of purely magnetic $\suinf$ black holes by Mavromatos and Winstanley \cite{mavromatos_infinitely_2000}.

Since the explicit process of deriving the gauge field ansatz and rewriting the field equations exists already \cite{mavromatos_infinitely_2000}, we merely quote the needed results here. As $N\rar\infty$, the matrices $A^{(N)}$-$D^{(N)}$ become the following series:
\begin{equation}\label{ansatzsumY}
\begin{split}
& A=\slim_{l=1}^\infty a_l(r)Y_{l,0}(\vartheta,\varphi), \quad B=\slim_{l=1}^\infty b_l(r)Y_{l,0}(\vartheta,\varphi),\\
& \\
& C=\slim_{l=1}^\infty c_l(r)Y_{l,1}(\vartheta,\varphi), \quad D=2P^0_1(\cos\vartheta),
\end{split}
\end{equation}
where $Y_{l,0}(\vartheta,\varphi)$, $Y_{l,\pm1}(\vartheta,\varphi)$ are spherical harmonics, given by
\begin{equation}\label{Ydefs}
\begin{split}
Y_{l,0}(\vartheta,\varphi)&=\left(\frac{2l+1}{4\pi}\right)^\frac{1}{2}P^0_l(\cos\vartheta),\\
Y_{l,\pm1}(\vartheta,\varphi)&=-\left(\frac{2l+1}{4\pi}\frac{(l-1)!}{(l+1)!}\right)^\frac{1}{2}e^{\pm i\varphi}P^1_l(\cos\vartheta),\\
\end{split}
\end{equation}
and $P^0_l(\cos\vartheta)$ and $P^1_l(\cos\vartheta)$ are Legendre functions of orders 0 and 1 respectively:
%
\begin{equation}\label{PL}
P^0_l(\cos\vartheta)=\frac{1}{2^l l!}\left(\frac{d}{d(\cos\vartheta)}\right)^l\left(\cos^2\vartheta-1\right)^l,
\end{equation}
and
\begin{equation}\label{PL1}
P^1_l(\cos\vartheta)=\left(1-\cos^2\vartheta\right)^\frac{1}{2}\left(\frac{d}{d(\cos\vartheta)}\right)P^0_l(\cos\vartheta).
\end{equation}
It should also be accepted that we assume the sums in \eqref{ansatzsumY} converge, which for instance implies that the functions $a_l$, $b_l$ and $c_l$ are uniformly bounded $\forall \,r,\,l$, and that the magnitudes of coefficient functions in \eqref{ansatzsumY} drop off sufficiently quickly for large $l$. This is necessary for the physicality of solutions.
%

\subsection{Dyonic field equations for AdS $\mk{su}(\infty)$ EYM theory}\label{suinflimit}

We begin with the well-known asymptotically AdS EYM field equations for $\sun$ \cite{baxter_soliton_2007}, in the form
\begin{equation}
G_{\mu\nu}+\Lambda g_{\mu\nu}=\,\kappa T_{\mu\nu},\qquad
\nabla_\lambda F^\lambda_{\,\,\mu}+[\mc{W}_\lambda,F^\lambda_{\,\,\mu}]=\,0,
\end{equation}
%
where in SI units $\kappa=8\pi G/c^4$. We are free to choose $\kappa$ by specifying convenient units, which we can do in order that the field equations agree with conventions taken in simpler cases; but we postpone this discussion for the time being. Here, $G_{\mu\nu}$ is the Einstein tensor, $\Lambda<0$ is the cosmological constant and the energy-momentum tensor is given by
\begin{equation}
T_{\mu\nu}=\mbox{Tr}\left[2g^{\rho\sigma}F_{\rho\mu}F_{\sigma\nu}-\frac{1}{2}g_{\mu\nu}F_{\rho\sigma}F^{\rho\sigma}\right],
\end{equation} 
where the antisymmetric field strength tensor is given by 
%
$F_{\mu\nu}=\partial_\mu\mc{W}_\nu-\partial_\nu\mc{W}_\mu+[\mc{W}_\mu,\mc{W}_\nu]$.
%
We use the signature $(-++\,+)$, with standard spherically symmetric `Schwarzschild-type' co-ordinates $(t,r,\theta,\phi)$, so that the metric has the form:
\begin{equation}\label{met}
ds^2=-\mu S^2dt^2+\mu^{-1}dr^2+r^2\left(d\theta^2+\sin^2\theta d\phi^2\right)
\end{equation}
where we are interested in static solutions, so that $\mu$ (the \emph{metric} function) and $S$ (the \emph{lapse} function) are both functions of $r$ alone, and $\mu$ can be expressed in the form
\begin{equation}\label{mu}
\mu(r)=1-\frac{2m(r)}{r}-\frac{\Lambda r^2}{3},
\end{equation} 
in which $m(r)$ is known as the \emph{mass function}. This reduces to the ordinary Schwarzschild-anti-de Sitter metric function when $m(r)$ is a constant. Note that in the case of black holes, $\mu(r)$ possesses one simple zero at the event horizon $r=r_h$, and we are interested only in exterior solutions, for which $r\geq r_h$. We can use knowledge of the $\sun$ system to simplify the equations slightly. There, the non-zero entries of the matrix $C$ are given by
\begin{equation}
C^{(N)}_{j,j+1}(r)=\omega_j(r) e^{i\nu_j(r)}
\end{equation}
for $2N-2$ real functions $\omega_j(r)$ and $\nu_j(r)$. Also, one of the Yang-Mills equations gives us $\frac{1}{2}\left(B^{(N)}_{jj}-B^{(N)}_{j+1,j+1}\right)+\nu^{\prime}_j=0$ (where from now on, $^\prime$ represents $d/dr$), so we fix our gauge such that $B^{(N)}\equiv0$, which implies that $\nu^{\prime}_j=0\,\,\forall j$ and therefore $\nu_j$ is constant $\forall j$. It can finally be shown that we can set $\nu_j=0\,\,\forall j$ due to the internal symmetry of the field equations 
%
$C^{(N)}\mapsto C^{(N)}e^{i\chi}$ 
%
for $\chi$ a real constant. We also emphasise that unlike \cite{mavromatos_infinitely_2000}, we are considering dyonic solutions, and so we need $A$ to be non-zero in general.

In order to take the limit $N\rar\infty$, we make the necessary substitution \cite{floratos_note_1989}
\begin{equation}\label{WNW}
\mc{W}\mapsto N\mc{W}.
\end{equation} 
Then the field equations can be derived as follows. The remaining Yang-Mills equations are
\begin{equation}
\begin{split}
0&= r^2\mu A^{\prime\prime}+r^2\mu\left(\frac{2}{r}-\frac{S^\prime}{S}\right)A^\prime-N^2[C,[A,C^\dagger]],\\
0&= r^2\mu C^{\prime\prime}+\left(2m-\frac{\kappa P}{r}-\frac{2\Lambda r^3}{3}-\frac{\kappa r^3\eta}{4S^2}\right)C^\prime+\frac{N^2}{\mu S^2}[A,[A,C]]+C+\frac{N^2}{2}[C,[C,C^\dagger]].\\
\end{split}
\end{equation}
The Einstein equations are given by
\begin{equation}\label{EEs}
m^\prime=\frac{\kappa}{2}\left(\frac{r^2\eta}{4S^2}+\frac{\zeta}{4\mu S^2}+\mu G+\frac{P}{r^2}\right),\qquad
\frac{S^\prime}{S}=\frac{2\kappa}{3}\left(\frac{G}{r}+\frac{\zeta}{4\mu^2 S^2r}\right),
\end{equation}
with
%
\begin{equation}\label{quants}
\begin{split}
\eta=-4N^2\mbox{Tr}(A^{\prime 2}),\quad& \zeta=-4N^2\mbox{Tr}(N^2[A,C][A,C^\dagger]),\\ 
G=N^2\mbox{Tr}(C^\prime C^{\dagger\prime}),\quad& P=\frac{N^2}{4}\mbox{Tr}\left(D-N[C,C^\dagger]\right)^2.
\end{split}
\end{equation}

Now we rewrite the field equations using the appropriate correspondences \cite{floratos_note_1989,mavromatos_infinitely_2000}. 
%
%
%
%
%
Using units in which the gauge coupling constant $g=1$,
%
%
 we use the substitutions
\begin{equation}\label{Traceop}
\begin{split}
\mbox{Tr}\,\rar&\frac{1}{4\pi N^2}\int_{S^2}dV,\\
%
%
N[Q,R]\rar & i\{Q(r,\vartheta,\varphi),R(r,\vartheta,\varphi)\},
\end{split}
\end{equation}
where we define the Poisson bracket $\{Q,R\}$ as
\begin{equation}
\{Q,R\}=\frac{\partial Q}{\partial(\cos\vartheta)}\frac{\partial R}{\partial\varphi}-\frac{\partial R}{\partial(\cos\vartheta)}\frac{\partial Q}{\partial\varphi}.
\end{equation}

Under these replacements, the Yang-Mills field equations become
\begin{equation}
\begin{split}
0&=r^2\mu A^{\prime\prime}+r^2\mu\left(\frac{2}{r}-\frac{S^\prime}{S}\right)A^\prime+\{C,\{A,C^\dagger\}\},\\
0&=r^2\mu C^{\prime\prime}+\left(2m-\frac{\kappa P}{r}-\frac{2\Lambda r^3}{3}-\frac{\kappa r^3\eta}{4S^2}\right)C^\prime-\frac{1}{\mu S^2}\{A,\{A,C\}\}\\
+&\,\,C-\frac{1}{2}\{C,\{C,C^\dagger\}\}.\\
\end{split}
\end{equation}
%
%

In the $N\rar\infty$ limit, the matrices $A$ and $C$ become functions of the internal angular co-ordinates $(\vartheta,\varphi)$. We may therefore write the ansatz of the previous subsection \eqref{ansatzsumY} in the more explicit forms
\begin{equation}\label{ansatzsumxi}
A(r,\vartheta,\varphi)=\frac{i}{2}\alpha(r,\vartheta),\quad C(r,\vartheta,\varphi)=\omega(r,\vartheta)e^{i\varphi},\quad D=-2\cos\vartheta
\end{equation}
for real functions $\alpha$ and $\omega$.
%
%

Finally, we may express the field equations explicitly, using the variable $\xi=\cos\vartheta$. The Yang-Mills equations become
\begin{subequations}\label{YMs}
	\begin{alignat}{10}
	0&=r^2\mu\frac{\partial^2\alpha}{\partial r^2}+r^2\mu\left(\frac{2}{r}-\frac{S^\prime}{S}\right)\frac{\partial\alpha}{\partial r}+\frac{\partial}{\partial\xi}\left(\omega^2\frac{\partial\alpha}{\partial\xi}\right),\label{YMMa}\\
	0&=r^2\mu\frac{\partial^2\omega}{\partial r^2}+\left(2m-\frac{\kappa P}{r}-\frac{2\Lambda r^3}{3}-\frac{\kappa r^3\eta}{4S^2}\right)\frac{\partial\omega}{\partial r}+\frac{r^2\omega}{4\mu S^2}\left(\frac{\partial\alpha}{\partial\xi}\right)^2\label{YMEa}\\
	&\,\,\,\,\,+\frac{\omega}{2}\frac{\partial^2}{\partial\xi^2}\left(\omega^2+\xi^2\right),\nonumber
	\end{alignat}
\end{subequations}
and the Einstein equations are the same as in \eqref{EEs}, but we can write the quantities \eqref{quants} more explicitly as
\begin{subequations}\label{xiquants}
	\begin{alignat}{10}
	\eta&=\frac{1}{2}\int_{-1}^1\left(\frac{\partial\alpha}{\partial r}\right)^2d\xi,\label{etaint}\\
	\zeta&=\frac{1}{2}\int_{-1}^1\omega^2\left(\frac{\partial\alpha}{\partial\xi}\right)^2d\xi,\label{zetaint}\\
	G&=\frac{1}{2}\int_{-1}^1\left(\frac{\partial\omega}{\partial r}\right)^2d\xi,\label{Gint}\\
	P&=\frac{1}{8}\int_{-1}^1\left(\frac{\partial}{\partial\xi}\left(\omega^2+\xi^2\right)\right)^2d\xi.\label{Pint}
	\end{alignat}
\end{subequations}

Finally, we note some constraints placed on the variables. The tracelessness of the matrix $A$ in $\sun$ becomes the constraint
\begin{equation}\label{alphaint}
\int_{-1}^1\alpha(r,\xi)d\xi=0.
\end{equation}
%
Also, examining the general forms of $P^0_{l}(\xi)$ \eqref{PL} and $P^1_l(\xi)$ \eqref{PL1}, we should be able to express the electric field variable as a power series in $\xi$, and the magnetic field variable $\omega(r,\xi)$ as a power series in $\xi$ multiplied by a factor of $\sin\vartheta=\sqrt{1-\xi^2}$. Hence, the gauge functions are restricted to be of the following forms:
\begin{subequations}\label{Ansatze}
	\begin{alignat}{10}
	\quad\alpha(r,\xi)&=\slim_{j=1}^\infty\alpha_j(r)\xi^{j}\label{alphaxisum},\\
	\omega(r,\xi)&=\sqrt{1-\xi^2}\slim_{j=0}^\infty\omega_j(r)\xi^j\label{omegaxisum}.
	\end{alignat}
\end{subequations}

It may be observed that if we let $\alpha\equiv 0$, we recover purely magnetic $\suinf$ field equations \cite{mavromatos_infinitely_2000}. We also note that the derivation above involved taking the $\sun$ field equations in matrix form and then taking the limit $N\rar\infty$. It is wise to point out that this process should commute, in that if we begin by taking the limit $N\rar\infty$ on the original field equations and gauge potential, we should obtain the same field equations; and indeed noting that the commutator brackets and the trace are both linear operations, it can be checked that this is the case. 

We are considering exterior black hole solutions here, and therefore solutions to these equations will be defined on the semi-infinite strip $(r,\xi)\in[r_h,\infty)\times\mc{I}$, where $\mc{I}=[-1,1]$. Therefore we will now briefly review the relevant field equation symmetries, and the boundary conditions for solutions to the field equations.

\subsection{Symmetries of the field equations}\label{symm}

We first note that the field equations 
respect the three independent symmetries 
\begin{equation}\label{mirsym}
	\xi\mapsto-\xi,\quad \alpha(r,\xi)\mapsto-\alpha(r,\xi), \quad\omega(r,\xi)\mapsto-\omega(r,\xi).
\end{equation}
These symmetries are helpful in defining unique charges for solutions (Section \ref{Charges}). We also have the simultaneous scaling symmetry
\begin{equation}\label{scalesym}
	t\mapsto\tau^{-1} t,\quad S\mapsto\tau S,\quad \alpha\mapsto\tau\alpha, \quad\tau\in\R.
\end{equation}
We can rely on this latter symmetry to ensure that the asymptotic boundary condition for $S(r)$ (below) may be met, if necessary by performing an overall rescaling of $S$.

\subsection{Boundary conditions for $r=r_h$}\label{BCrh}

For convenience, we use the notation $f_h\equiv f(r_h)$ and $f^\prime_h\equiv\pd{f}{r}\Big|_{r=r_h}$. At $r=r_h$, we require a regular event horizon, meaning that $\mu_h=0$, so that \eqref{mu} implies that
\begin{equation}\label{mh}
m_h=\frac{r_h}{2}-\frac{\Lambda r_h^3}{6}.
\end{equation}
The ``free parameters'' for the gauge fields in this case, which will be functions of $\xi$ alone, can be found by using $\mu_h=0$ in the Yang-Mills equations \eqref{YMs}. Doing this shows that $\alpha(r_h,\xi)=0\,\,\,\forall\xi\in\mc{I}$. 
Then, the analysis shows that our solutions are entirely determined by the two functions $\omega_h(\xi)$ and $\alpha^\prime_h(\xi)$.
%
%
All higher order derivatives at the event horizon are determined from these: For instance, we have
\begin{equation}
\omega^\prime_h(\xi)=-\frac{\omega_h(\xi)\left(1+\frac{1}{2}\frac{d^2}{d\xi^2}\left(\omega^2_h(\xi)\right)\right)}{r_h^2\left(2m_h-\frac{\kappa P_h}{r_h}-\frac{2\Lambda r_h^3}{3}-\frac{\kappa r_h^3\eta_h}{4S_h^2}\right)},\quad
S^\prime_h=\frac{\kappa S_h}{r_h}\int_{-1}^1\left(\omega^\prime_h(\xi)\right)^2d\xi,
\end{equation}
where we note that $P_h$ depends on $\omega_h$, and $\eta_h$ is determined by $\alpha^\prime_{h}$. Therefore at this boundary we also have infinitely many parameters that need to be specified -- the expansion coefficients of the two functions $\omega_h(\xi)$ and $\alpha^\prime_h(\xi)$.

We should also finally note a weak constraint on $\omega$ and $\alpha^\prime$ near the horizon, given by the condition $\mu^\prime_h>0$ which ensures a non-extremal event horizon: setting $\mu_h=\alpha_{h}=0$ in the field equation for $m'$ in \eqref{EEs} gives us
\begin{equation}\label{mph}
2m'(r_h)=\kappa\left(\disfrac{r^2_h\eta_h}{4S_h^2}+\frac{P_h}{r_h^2}\right)<1-\Lambda r_h^2,
\end{equation}
%
which gives us an upper bound on the (negative) value of $\Lambda$:
\begin{equation}\label{lambdacon}
\Lambda<\frac{1}{r_h^2}-\kappa\left(\frac{\eta_h}{4S_h^2}+\frac{P_h}{r_h^4}\right).
\end{equation}
It is worth remarking that this implies that $|\Lambda|$ may take arbitrarily small values 
if the gauge fields are sufficiently small, which is a necessary condition if these models are to be candidates for `regulator models' for asymptotically flat solutions -- see the discussion at the end of Section \ref{SGF}.

\subsection{Boundary conditions as $r\rar\infty$}\label{BCinf}

We change variables to $z=r^{-1}$ and expand the field variables in power series about $z=0$, to investigate the asymptotic boundary of the AdS space. We find using the field equations \eqref{EEs}, \eqref{YMs} that
\begin{equation}\label{infexp}
\begin{split}
m(z)&=M+m_1z+O(z^2),\\
S(z)&=1+S_4z^4+O(z^5),\\
\alpha(z,\xi)&=\alpha_\infty(\xi)+\mc{A}(\xi)z+O(z^2),\\
\omega(z,\xi)&=\omega_\infty(\xi)+\mc{W}(\xi)z+O(z^2),\\
\end{split}
\end{equation}
where
\begin{equation}\label{m1}
\begin{split}
m_1&=-\frac{\kappa}{2}\left(\frac{1}{8}\int\limits_{-1}^1\mc{A}^2d\xi+\frac{1}{8}\int\limits_{-1}^1\left(\pd{}{\xi}\left(\omega_\infty^2+\xi^2\right)\right)^2d\xi-\frac{3}{8\Lambda}\int\limits_{-1}^1\omega_\infty^2\left(\pd{\alpha_\infty}{\xi}\right)^2d\xi-
\frac{\Lambda}{6}\int\limits_{-1}^1\mc{W}^2d\xi\right),\\
S_4&=-\frac{\kappa}{2}\left(\frac{3}{8\Lambda^2}\int_{-1}^1\omega_\infty^2\left(\pd{\alpha_\infty}{\xi}\right)^2d\xi+\frac{1}{6}\int_{-1}^1\mc{W}^2d\xi\right).
\end{split}
\end{equation}
There are no constraints on the first two parameters of each gauge field, with higher order terms determined entirely by those parameters. Therefore the equations at $z=0$ (i.e. as $r\rar\infty$) are fixed by four functions of $\xi$ -- $\omega_\infty(\xi)$, $\alpha_\infty(\xi)$, $\mc{W}(\xi)=-\lim_{r\rar\infty}r^2\pd{\omega}{r}$, $\mc{A}(\xi)=-\lim_{r\rar\infty}r^2\pd{\alpha}{r}$ --  and the constant ADM mass $M$.

\subsection{Boundary conditions for $\xi=\pm1$}

Firstly, we note that all spherical harmonics (except those for $m=0$) vanish at $\theta=0$ or $\pi$, corresponding to $\xi=\pm1$. This means we can define $\omega$ on the $\xi$-boundaries as
\begin{equation}\label{omegaxibcs}
\omega(r,\pm1)\equiv 0\,\,\forall r,
\end{equation}
which is implied by the ans\"{a}tze \eqref{Ansatze}. 

As for the electric field, we can see that $A$ is based on the $m=0$ harmonics, so it does not vanish on the $\xi$-boundaries, since $P_l(\pm 1)=(\pm 1)^l$. Therefore, the $\xi$ boundary conditions for $\alpha$ 
in the general case are a little unclear. However, 
these boundary conditions do not come up explicitly in our investigations, so we will simply assume that the electric field is regular $\forall\,\xi\in\mc{I}$, which would be necessary anyway for physicality. 

\section{Embedded `trivial' solutions}\label{Embed}

We may find the following trivial solutions to the field equations, for which existence can be proven, with embeddings given as follows.

\subsection{The embedded $\mk{su}(2)$ solution}\label{Embedsu2}

We expect $\suinf$ field equations to be able to be embedded in $\essu$ since $\essu$ is a subalgebra of $\suinf$. Hence, we let 
\begin{equation}\label{su2embed}
\alpha(r,\xi)=2\xi\alpha_*(r),\quad 
\mbox{and}\quad\omega(r,\xi)=\omega_*(r)\sqrt{1-\xi^2}.
\end{equation}
The Yang-Mills equations become
\begin{equation}
\begin{split}
0&=-r^2\mu\alpha_*^{\prime\prime}+r^2\mu\left(\frac{S^\prime}{S}-\frac{2}{r}\right)\alpha_*^\prime+2\omega_*^2\alpha_*,\\
0&=r^2\mu\omega_*^{\prime\prime}+\left(2m-\frac{\kappa P}{r}-\frac{2\Lambda r^3}{3}-\frac{\kappa r^3\eta}{4S^2}\right)\omega_*^\prime+\frac{r^2\omega_*\alpha_*^2}{\mu S^2}+\omega_*(1-\omega_*^2),
\end{split}
\end{equation}
and the 
%
%
Einstein equations become
\begin{equation}
\begin{split}
m^\prime(r)&=\frac{\kappa}{3}\left(\frac{r^2\alpha_*^{\prime 2}}{2S^2}+\frac{\omega_*^2\alpha_*^2}{\mu S^2}+\mu\omega_*^{\prime 2}+\frac{1}{2r^2}(1-\omega_*^2)^2\right),\\
\frac{S^\prime}{S}&=\frac{2\kappa}{3}\left(\frac{\omega_*^{\prime 2}}{r}+\frac{\omega_*^2\alpha_*^2}{\mu^2S^2r}\right).\\
\end{split}
\end{equation}

These are exactly the $\essu$ EYM field equations, for which solutions are known to exist which are stable under linear perturbations \cite{winstanley_existence_1999} and can be characterised by one global charge \cite{shepherd_characterizing_2012}. 

As in \cite{mavromatos_infinitely_2000}, we note a slightly odd thing -- working with finite $N$, it is usual to take units where $c=1$ and $4\pi G=1$, so that $\kappa=2$, which in turn means that the $\essu$ solution appears as an embedding. However, in $\suinf$ we find that in order for the $\mk{su}(2)$ solution to appear as an embedding, we need $\kappa=3$, corresponding to the choice $c=1$, $4\pi G=3/2$, which we will do from here on unless otherwise stated. The reason for this was not discussed in \cite{mavromatos_infinitely_2000}, but we offer a suggestion. In \cite{floratos_note_1989}, we can see that the limit $N\rar\infty$ rescales the gauge coupling constant $g_N$ for $\sun$ to another finite value $g$, with $\lim_{N\rar\infty}N^{-1}g_N=g$. The only place that $\kappa$ appears in the field equations are in the Einstein equations \eqref{EEs}, and they are always attached to a trace operation, which would carry factors of $g_N$. Therefore, it is possible that $\kappa$ also experiences some kind of rescaling in the limit $N\rar\infty$. This is just a reasonable possibility; exploring this issue in detail would go beyond the scope of this paper, but seems worthy of future investigation.


\subsection{The Schwarzschild-Anti-de Sitter solution (SAdS)}\label{EmbedSAdS}

Here, we let $\alpha(r,\xi)\equiv 0$ and $\omega(r,\xi)\equiv\sqrt{1-\xi^2}$. We note that the electric and magnetic charges are given by $\mc{Q}^2_E=\mc{Q}^2_M=0$. Furthermore, we find $m(r)\equiv M$, the (constant) ADM mass of the solution; and $S^\prime=0$, so we can let $S\equiv 1$ to comply with the asymptotic boundary conditions. If in addition we let $M=0$, we have pure AdS space as a vacuum solution.

\subsection{The Reissner-N\"{o}rdstrom-Anti-de Sitter solution (RNAdS)}\label{EmbedRNAdS}

Here, we let $\alpha(r,\xi)\equiv 0$ and $\omega(r,\xi)\equiv0$. The Yang-Mills equations are identically satisfied, and it is clear that $\eta=G=\zeta=0$, so that again $S^\prime=0$ and thus $S\equiv 1$. Also we find that $P=1/3$. Hence the mass of the solution $m(r)$ is
\begin{equation}\label{mRNAdS}
m(r)=M-\frac{1}{2r},
\end{equation}
and so
\begin{equation}
\mu(r)=1-\frac{2M}{r}-\frac{\Lambda r^2}{3}+\frac{\mc{Q}_M^2}{r^2}
\end{equation}
where $\mc{Q}_M^2=1$, and $\mc{Q}_E^2=0$.

\subsection{Purely Abelian solutions}\label{PureAb}

In the cases of $\essu$ \cite{nolan_existence_2012}, $\sun$ \cite{baxter_existence_2016} and for a general semisimple gauge group \cite{baxter_existence_2018}, we find that there are so-called \emph{purely Abelian} solutions for which the magnetic gauge field vanishes, and the electric gauge field does not. We find an analogue of those solutions here.
We shall let $\omega$ be identically zero and let $\alpha$ take the form of a Coulomb-type potential:
\begin{equation}\label{pureabsol}
\alpha(r,\xi)\equiv\frac{\psi(\xi)}{r},\quad \omega(r,\xi)\equiv 0,
\end{equation}
for some arbitrary function $\psi(\xi)$, constrained by \eqref{alphaint} with 
\begin{equation}\label{intcons}
\int_{-1}^1\psi(\xi)d\xi=0.
\end{equation} 
%
Then both Yang-Mills equations are satisfied automatically for \eqref{pureabsol}. 
%
%
The Einstein equations give $S(r)\equiv1$, and
\begin{equation}\label{PureAbM}
m(r)=M-\frac{1}{2r}\left(\frac{3}{8}\int_{-1}^1\psi^2(\xi)d\xi+1\right).
\end{equation}
We can manipulate this to obtain
\begin{equation}
\mu(r)=1-\frac{2M}{r}-\frac{\Lambda r^2}{3}+\frac{\mc{Q}^2}{r^2},
\end{equation}
where we have defined the total squared charge 
%
$\mc{Q}^2=\mc{Q}_E^2+\mc{Q}_M^2$, 
%
for the magnetic charge $\mc{Q}_M^2=1$, and the electric charge
\begin{equation}\label{QE2}
\mc{Q}^2_E=\frac{3}{8}\int_{-1}^1\psi^2(\xi)d\xi.
\end{equation}
This looks like an odd value for the charge, but substituting in $\psi(\xi)=2\psi_0\xi$ (which obeys \eqref{intcons}) gives us $\mc{Q}_E^2=\psi_0^2$ in agreement with the $\essu$ case \cite{nolan_existence_2012}. 
%
%

From \eqref{QE2} it naively seems as if only one constant is needed to specify the solution: the value of $\mc{Q}_E^2$. However, we can write $\psi$ in the form \eqref{Ansatze},
\begin{equation}
\psi(\xi)=\slim_{j=1}^\infty\psi_j\xi^{j},
\end{equation}
and then it is easy to see that \eqref{pureabsol} is specified by an infinitely many arbitrary parameters. Hence this solution represents an infinite product of trivial $\mk{u}(1)$ embedded solutions.

\subsection{Re-obtaining the field equations for $\sun$: An analogy}\label{Embedsun}

Extracting the $\sun$ system back from the $\suinf$ equations is surprisingly elementary. Essentially we `re-discretise' our equations in the variable $\xi$, using the `method of lines'. 
%
We imagine dividing the range $\xi\in[-1,1]$ into a finite number of discrete lines such that
\begin{equation}
\xi_j=-1+\frac{2j}{N},\,\,j\in\{1,2,...,N-1\},
\end{equation}
and we use finite difference approximations on the field equations, which will result in $N-1$ ODEs. All functions are rewritten as $\hat{f}_j(r)\equiv f(r,\xi_j)$. Derivatives of $\omega$ are written as
\begin{equation}\label{derw}
\begin{split}
\frac{\partial \omega}{\partial\xi}&\rar\frac{N}{2}\left(\hat{\omega}_j-\hat{\omega}_{j-1}\right),\\
\frac{\partial^2 \omega}{\partial\xi^2}&\rar\frac{N^2}{4}\left(\hat{\omega}_{j+1}-2\hat{\omega}_j+\hat{\omega}_{j-1}\right),\\
\end{split}
\end{equation}
and (for purely notational reasons) derivatives of $\alpha$ are defined as the 
%
%
%
%
\emph{forward} difference between the $\alpha_j$, i.e.
\begin{equation}
\pd{\alpha}{\xi}\rar-\frac{N}{2}\left(\hat{\alpha}_{j}-\hat{\alpha}_{j+1}\right),\,\,0<j\leq N-1. 
\end{equation}
%
%
%
We approximate all integrals with the rectangle rule as
\begin{equation}
\int\limits_{-1}^1fd\xi\rar\frac{2}{N}\slim_{j=1}^{N}\hat{f}_j(r). 
\end{equation} 
%
Finally, we rescale variables as follows:
%
%
%
%
%
\begin{equation}
\begin{array}{llll}
\quad\bar{N}\equiv\disfrac{2}{N^{3/2}},&\quad\hat{\omega}_j=\disfrac{2}{N}\bar{\omega}_j,&\quad\hat{\alpha}_j=\disfrac{2}{N}\bar{\alpha}_j,& \\[10pt]
\quad\bar{r}=\bar{N}r,&\quad\bar{m}=\bar{N}m,&\quad\bar{\Lambda}=\bar{N}^{-2}\Lambda, &\quad\bar{S}=\bar{N}S.\\
\end{array}
\end{equation}

We first note that the form of $\mu$ is unchanged:
\begin{equation}
\mu(\bar{r})=1-\frac{2\bar{m}}{\bar{r}}+\frac{\bar{\Lambda}\bar{r}^2}{3}.
\end{equation}
The Einstein equations become
\begin{subequations}\label{SUNEE}
	\begin{alignat}{10}
	\frac{d\bar{m}}{d\bar{r}}&=\frac{\kappa}{2}\left(\frac{\bar{r}^2\bar{\eta}}{4\bar{S}^2}+\frac{\bar{\zeta}}{4\mu \bar{S}^2}+\mu \bar{G}+\frac{\bar{P}}{\bar{r}^2}\right),\label{E1}\\
	\frac{1}{\bar{S}}\frac{d\bar{S}}{d\bar{r}}&=\kappa\left(\frac{\bar{G}}{\bar{r}}+\frac{\bar{\zeta}}{4\mu^2 \bar{S}^2\bar{r}}\right),\label{E2}
	\end{alignat}
\end{subequations}
with
%
\begin{eqnarray}
\bar{\eta}=\slim_{j=1}^N\left(\disfrac{d\bar{\alpha}_j}{d\bar{r}}\right)^2, & 
\quad\bar{\zeta}=\slim_{j=1}^{N-1}\bar{\omega}_j^2\left(\bar{\alpha}_j-\bar{\alpha}_{j+1}\right)^2,\\
\bar{G}=\slim_{j=1}^{N-1}\left(\disfrac{d\bar{\omega}_j}{d\bar{r}}\right)^2, & 
\quad\bar{P}=\disfrac{1}{4}\slim_{j=1}^N\left(\bar{\omega}_j^2-\bar{\omega}_{j-1}^2-N-1+2j\right)^2;
\end{eqnarray}
%
and the Yang-Mills equations become
\begin{equation}\label{SUNYM}
\begin{split}
0&=	\bar{r}^2\mu\disfrac{d^2\bar{\alpha}_j}{d\bar{r}^2}+\bar{r}^2\left(\disfrac{2}{\bar{r}}-\disfrac{1}{\bar{S}}\disfrac{d\bar{S}}{d\bar{r}}\right)\disfrac{d\bar{\alpha}_j}{d\bar{r}}+\bar{\omega}^2_{j}\left(\bar{\alpha}_{j}-\bar{\alpha}_{j+1}\right)-\bar{\omega}_{j-1}^2\left(\bar{\alpha}_{j-1}-\bar{\alpha}_j\right).\\
0&=r^2\mu\disfrac{d^2\bar{\omega}_j}{d\bar{r}^2}+r^2\left(\disfrac{d\mu}{d\bar{r}}+\disfrac{\mu}{\bar{S}}\disfrac{d\bar{S}}{d\bar{r}}\right)\disfrac{d\bar{\omega}_j}{d\bar{r}}+\disfrac{\bar{r}^2}{\mu \bar{S}^2}\bar{\omega}_j\left(\bar{\alpha}_j-\bar{\alpha}_{j+1}\right)^2\\
&+\bar{\omega}_j\left(1-\bar{\omega}_j^2+\disfrac{1}{2}\left(\bar{\omega}_{j-1}^2+\bar{\omega}_{j+1}^2\right)\right).
\end{split}
\end{equation}

If we here use conventional units for $\sun$ in which we set $c=4\pi G=1$ and hence $\kappa=2$, then it is immediately seen that these are the field equations for the spherically symmetric $\sun$ EYM system, solutions to which have been proven in several regimes \cite{baxter_existence_2016}.

The fact that it is so easy to reduce the $\suinf$ equations back down to the $\sun$ equations with a na\"{i}ve finite difference scheme is somewhat surprising. It should be emphasised that using this method on such non-linear, non-globally-hyperbolic equations would not in general converge as a numerical scheme; but all we wish to show is that this method reproduces the field equations, and should be more viewed as a `dictionary of correspondences' in the same sense as was required to take the limit $N\rar\infty$ in Section \ref{suinflimit} (particularly Equations \eqref{WNW} and \eqref{Traceop}). In any case, we already know that analytical solutions exist to Equations \eqref{SUNEE} and \eqref{SUNYM} for all finite values of $N$ \cite{baxter_existence_2016}, and many numerical solutions have been found (see e.g.  \cite{baxter_existence_2008,nolan_existence_2012}) so this need not concern us. That this correspondence works is largely due the fact that the Cartan matrix for $\sun$, applied as a linear transform on a list of functions $\{\hat{f}_j\}$, results in a vector of terms that resemble 2nd derivatives written in a central finite difference scheme \eqref{derw}. It thus seems like a coincidence particular to this case. 

It does hint though that expressions for important quantities to $\suinf$ black holes can in principle be quickly converted into analogous expressions for $\sun$ black holes, which means that $\suinf$ could be a good testing ground for obtaining or checking results in $\sun$, one which uses 4 partial integro-differential equations instead of $2N$ ODEs, and may be amenable to a different class of analytical methods.

\section{Solutions with small gauge fields}\label{SGF}

The first kind of non-trivial solution we search for are solutions where both gauge fields are small. This will linearise the equations to zeroeth order, making them easier to solve, but it is also of interest for AdS/CFT, since it has previously been proven that the correspondence requires the gauge fields involved be small \cite{graham_einstein_1991}. Given that we recover the extremal RNAdS solution if $\omega=\alpha=0$, we are essentially considering solutions in some neighbourhood of these.

We let $\epsilon>0$ be some sufficiently small parameter, and assume we have gauge fields such that $|\alpha(r,\xi)|\leq\epsilon$, $|\omega(r,\xi)|\leq\epsilon$ $\forall r$. We use the expansions
\begin{equation}
\begin{split}
m(r)&=\check{m}_0(r)+\epsilon \check{m}_1(r)+O(\epsilon^2),\\
S(r)&=\check{S}_0(r)+\epsilon \check{S}_1(r)+O(\epsilon^2),\\
\alpha(r,\xi)&=\epsilon\check{\alpha}_0(r,\xi)+
O(\epsilon^2),\\
\omega(r,\xi)&=\epsilon\check{\omega}_0(r,\xi)+O(\epsilon^2).
\end{split}
\end{equation}
Substituting these expansions into the field equations, we easily find that $\check{m}_1=\check{S}_1=0$. 

At zeroeth order, we find that $\check{S}^\prime_0(r)=0$, since $G=O(\epsilon^2)$ and $\zeta=O(\epsilon^4)$, and so we set $\check{S}_0\equiv 1$ for the correct asymptotic limit \eqref{infexp}. Also, $\eta=O(\epsilon^2)$ and $P=1/3+O(\epsilon^2)$, so we obtain
\begin{equation}\label{SGF_mgen}
\check{m}^\prime_0(r)=\frac{1}{r^2}\,\,\implies\,\,\check{m}_0(r)=\mc{M}-\frac{1}{2r}
\end{equation}
for some arbitrary constant $\mc{M}$. Using boundary conditions at the event horizon, we obtain
\begin{equation}\label{SGF_m}
\check{m}_0(r)=\frac{r_h}{2}-\frac{\Lambda r_h^3}{6}+\frac{1}{2r_h}-\frac{1}{2r}.
\end{equation}
This is identical to the situation in \cite{mavromatos_infinitely_2000}, and it was shown there that this is a good approximation to the mass function for $\essu$ solutions. 

The Yang-Mills equations become, to zeroeth order,
\begin{equation}\label{SGF_YM}
\begin{split}
0&=\frac{\partial}{\partial r}\left(r^2\frac{\partial\check{\alpha}_0}{\partial r}\right),\\
0&=\left(1-\frac{2\check{m}_0}{r}-\frac{\Lambda r^2}{3}\right)\frac{\partial^2\check{\omega}_0}{\partial r^2}+\left(\frac{2\check{m}_0}{r^2}-\frac{2\Lambda r}{3}-\frac{1}{r^3}\right)\frac{\partial\check{\omega}_0}{\partial r}+\frac{\check{\omega}_0}{r^2}.\\
\end{split}
\end{equation}
%
Since these equations have only $r$ derivatives, the $\xi$ dependence is arbitrary. 
The magnetic gauge field equation is exactly the purely magnetic $\essu$ Yang-Mills equation with small $\omega$, multiplied by some arbitrary $\xi$-dependence, to which solutions exist \cite{winstanley_existence_1999, mavromatos_infinitely_2000}. Also, the electric equation can be directly solved, giving
\begin{equation}\label{asol}
\check{\alpha}_0(r,\xi)=\left(\frac{1}{r_h}-\frac{1}{r}\right)\mc{Z}(\xi)+\mc{Y}(\xi).
\end{equation}
Using the boundary condition $\alpha(r_h,\xi)=0$, we find we must take $\mc{Y}(\xi)\equiv0$; and using the asymptotic boundary conditions, we can fix $\mc{Z}(\xi)=-\mc{A}(\xi)$ \eqref{infexp}. Hence we obtain
\begin{equation}\label{SGF_alpha}
\check{\alpha}_0(r,\xi)=\mc{A}(\xi)\left(\frac{1}{r}-\frac{1}{r_h}\right).
\end{equation}
%
It is worth noting that the above expression looks remarkably similar to the expression for $A^\perp$ in the analysis of black holes in EYM theories for general semisimple gauge groups, if we there let $S\equiv 1$ \cite{baxter_existence_2018}.

Finally we use the ans\"{a}tze \eqref{Ansatze} to write the solutions out explicitly:
\begin{equation}\label{omaleps}
\begin{split}
\alpha(r,\xi)=&\left(\left(\frac{1}{r}-\frac{1}{r_h}\right)\slim_{j=0}^\infty \mc{A}_j\xi^{j}\right)\epsilon+O(\epsilon^2),\\
\omega(r,\xi)=&\left(\omega_*(r)\sqrt{1-\xi^2}\slim_{j=0}^\infty \mc{U}_j\xi^j\right)\epsilon+O(\epsilon^2).
\end{split}
\end{equation}
for two infinite sets of constants $\{\mc{A}_j,\mc{U}_j\}$, where $\mc{A}_j$ are the expansion coefficients of $\mc{A}(\xi)$ \eqref{infexp}, $\mc{U}_j$ are the expansion coefficients of an arbitrary function of $\xi$, and $\omega_*(r)$ is the magnetic field variable for a \emph{small} purely magnetic $\essu$ solution. This manifestly shows that the solution is determined by infinitely many gauge field parameters.

It should be noted that these solutions do not exist if $\Lambda=0$. There, any solution which begins small at the event horizon can in general grow arbitrarily large (though finite) asymptotically. However here, we find that solutions with small gauge fields are valid for all values of $\Lambda<0$, even arbitrarily small, and remain small asymptotically, i.e. at the boundary of the AdS space. Hence, these solutions expand on those in \cite{mavromatos_infinitely_2000} which when $\Lambda\rar0^-$, were referred to as \emph{regulator} models in the purely magnetic case, in the sense that we may obtain good results for asymptotically flat space, which may in some cases be motivated by relevant phenomenology.

\section{Solutions as $|\Lambda|\rar\infty$}\label{LargeLambda}

Whenever solutions to asymptotically AdS EYM theories are discussed, an investigation into the regime where $|\Lambda|$ is large cannot be far away, and the motivation for this has been multifold, which we can illustrate by referring to the case of purely magnetic $\sun$ solutions. In numerical simulations, we there find that the regions of the initial parameter space admitting solutions shrinks as $N$ grows, but in the limit of large $|\Lambda|$, we find that the \emph{entire} parameter space of initial values will produce regular solutions \cite{baxter_soliton_2007}. In analytical investigations, this limit has been fruitful in allowing us to prove the existence of solutions, and is necessary for the stability of solutions and the definition of uniquely characterising charges \cite{baxter_existence_2008,baxter_stability_2016,shepherd_characterizing_2012}. In addition, there are motivations from black hole thermodynamics: the regime corresponds to large black holes, possessing a stable Hartle-Hawking state, which is of relevance to the question of information loss during Hawking evaporation \cite{hawking_thermodynamics_1983}. It is also this regime where string corrections become negligible in the bulk and the field theory approximation may be used \cite{maldacena_large_1998,witten_anti-sitter_1998}.

Our strategy will be to assume power series expansions for the field variables essentially in the parameter $\Lambda^{-1}$, and therefore these will be asymptotic expansions -- for a large enough value of $|\Lambda|$, truncating the series expansions will give an approximate solution to the field equations which becomes more exact the more terms we take in the truncation. Given our physicality assumption that the infinite sums in \eqref{ansatzsumY} converge, this implies the convergence of the sums in \eqref{Ansatze}, which further implies that the coefficient functions in our large-$|\Lambda|$ expansions will be uniformly bounded. Therefore, the full series will be an exact solution. 

We should note that this analysis will differ from a similar one in \cite{mavromatos_infinitely_2000}, because in anticipation of Section \ref{Charges}, we want to use \emph{asymptotic} boundary conditions to specify the gauge fields, rather than the boundary at the event horizon $r=r_h$. For $m$ and $S$ we will see that we are free to use a combination of the conditions, and the regularity of the event horizon, to fix those functions. The analysis would be simple enough to repeat using boundary conditions at $r=r_h$, and would just complicate the expressions given at the end of Section \ref{Charges}.

First, we let
\begin{equation}\label{lamdef}
\lambda\equiv\frac{1}{\Lambda},\qquad \tilde{m}(r)=\lambda m(r),
\end{equation}
and consider the regime of $\lambda$ very small. Writing the field equations \eqref{EEs} and \eqref{YMs} in these new variables and letting $\lambda=0$, we find the resulting equations:
\begin{equation}
\tilde{m}^\prime=0,\quad S^\prime=0,\quad \pd{\alpha}{r}=0,\quad \pd{\omega}{r}=0.
\end{equation}
We can easily solve the resulting equations by integrating, and then impose boundary conditions, at $r=r_h$ for $m$ and $\alpha$ and as $r\rar\infty$ for $S$ (see Sections \ref{BCrh} and \ref{BCinf}), to fix most of the constants and functions of integration. Hence we obtain the solution
\begin{equation}\label{lam0}
\tilde{m}(r)=-\frac{r_h^3}{6}, \quad S(r)=1,\quad
\alpha(r,\xi)=0, \quad \omega(r,\xi)=\mc{K}(\xi),
\end{equation}
which manifestly is uniquely determined by $r_h$ and $\mc{K}(\xi)$, some arbitrary function of $\xi$. Furthermore, we will impose asymptotic boundary conditions on $\omega$, and let $\mc{K}(\xi)=\omega_\infty(\xi)$, in anticipation of Section \ref{Charges}. 

It must be emphasised that we \emph{cannot} expect $\lambda=0$ to give us a sensible solution to the field equations, for several reasons: it would mean that the AdS radius of curvature $\ell=\sqrt{\frac{-3}{\Lambda}}$ would be zero, so $\mu$ would become singular, and in addition the definitions \eqref{lamdef} break down. However, we can investigate the case where $\lambda$ is arbitrarily close to 0, to find solutions in some $\lambda$-neighbourhood of \eqref{lam0}. Hence, our aim here is to find solutions by assuming asymptotic expansions on the field variables as power series in $\lambda$, for a fixed value of $\lambda$, which are well-defined in a neighbourhood of $\lambda=0$.

Regarding the electric field variable $\alpha$, we have found in previous treatments that establishing existence as $|\Lambda|\rar\infty$ for dyonic solutions was difficult unless we took the electric gauge field as being small. Hence here, noting \eqref{lam0}, and recalling that we are searching for solutions in some neighbourhood of $\lambda=0$, we let the electric field be $O(\lambda)$, so as $\lambda$ becomes very small, so will $\alpha$.

Therefore, motivated by \eqref{lam0} and the earlier discussion, we define the following expansions for $\lambda$ small:
\begin{equation}\label{laminfexp}
\begin{array}{lll}
S(r)=1+\slim_{j=1}^\infty \tilde{S}_j(r)\lambda^j, & \quad &	\tilde{m}(r)=\frac{-r_h^3}{6}+\slim_{j=1}^\infty\tilde{m}_j(r)\lambda^j,\\
\alpha(r,\xi)=\slim_{j=1}^\infty\tilde{\alpha}_j(r,\xi)\lambda^j, &  \quad & \omega(r,\xi)=\omega_\infty(\xi)+\slim_{j=1}^\infty\tilde{\omega}_j(r,\xi)\lambda^j.\\
\end{array}
\end{equation}
The zeroeth order terms in $\lambda$ confirm \eqref{lam0}. To first order, we find that
\begin{equation}\label{O1}
\begin{split}
\tilde{S}_1^\prime&=0,\\
\tilde{m}_1^\prime&=\frac{K_1}{2r^2},\\
0&=r^2\pd{^2\tilde{\alpha}_1}{r^2}+2r\frac{\partial\tilde{\alpha}_1}{\partial r},\\		
\mc{F}(\xi)&=r^2\left(r^2-\frac{r_h^3}{r}\right)\frac{\partial^2\tilde{\omega}_1}{\partial 		r^2}+(2r^3+r_h^3)\frac{\partial\tilde{\omega}_1}{\partial r},\\
\end{split}
\end{equation}
where
\begin{equation}\label{F1andK1}
K_1=\frac{3}{8}\int_{-1}^1\left(\frac{\partial}{\partial\xi}\left(\omega_\infty^2+\xi^2\right)\right)^2d\xi,\quad
\mc{F}(\xi)=\frac{3\omega_\infty}{2}\pd{^2}{\xi^2}\left(\omega_\infty^2+\xi^2\right).
\end{equation}
In order to solve Equations \eqref{O1}, and indeed, to solve the field equations to all higher orders in $\lambda$, we will need some boundary conditions consistent with our choices of arbitrary constants and functions at zeroeth order \eqref{lam0} and with the boundary conditions given in Sections \ref{BCrh} and \ref{BCinf}. We therefore choose the following conditions:
\begin{equation}\label{LLBCs}
\begin{split}
&\tilde{m}_1(r_h)=\frac{r_h}{2},\quad\tilde{m}_j(r_h)=0\mbox{  for } j>1;\\
&\lim_{r\rar\infty}\tilde{S}_j=0\mbox{  for } j\geq1;\\
&\tilde{\alpha}_j(r_h,\xi)=0\mbox{  for } j\geq1,\\
&\lim_{r\rar\infty}r^2\pd{\tilde{\alpha}_1}{r}=-\mc{A}(\xi),\quad\lim_{r\rar\infty}r^2\pd{\tilde{\alpha}_j}{r}=0\mbox{  for } j>1;\\
&\lim_{r\rar\infty}\tilde{\omega}_j=0\mbox{  for } j\geq1;\\		
\end{split}
\end{equation}
where $\mc{A}$ is one of the electric asymptotic boundary functions (see Section \eqref{BCinf}).

Noting these, we find from \eqref{O1} that $\tilde{S}_1=0$. Also, $\tilde{m}_1$ is given by
\begin{equation}\label{q11}
\tilde{m}_1(r)=K_2-\frac{K_1}{2r},
\end{equation}
for an arbitrary constant $K_2$, which may be fixed in terms of $K_1$ by using \eqref{LLBCs}, to give
\begin{equation}\label{q1}
\tilde{m}_1(r)=\frac{r_h}{2}+\frac{K_1}{2}\left(\frac{1}{r_h}-\frac{1}{r}\right).
\end{equation}
It can thus be seen that up to first order, the geometry resembles RNAdS \eqref{mRNAdS}, as expected since at this order the gauge fields are both small -- compare Equation \eqref{SGF_m}. In fact it will later be seen that $K_1$ can be interpreted as the total magnetic charge. 

We find that $\tilde{\alpha}_1$ is given by
\begin{equation}\label{E12}
\tilde{\alpha}_1(r,\xi)=\mc{J}(\xi)-\frac{\mc{H}(\xi)}{r},
\end{equation}
for two arbitrary functions $\mc{H}(\xi)$ and $\mc{J}(\xi)$. Given that $\alpha(r_h,\xi)=0$ \eqref{LLBCs}, we must have $\mc{H}=r_h\mc{J}$ and so
\begin{equation}\label{E_1H}
\tilde{\alpha}_1(r,\xi)=\left(\frac{1}{r_h}-\frac{1}{r}\right)\mc{H}(\xi).
\end{equation}
In addition, using $\lim_{r\rar\infty}r^2\pd{\tilde{\alpha}_1}{r}=-\mc{A}(\xi)$, we can fix the function of integration as  $\mc{H}(\xi)=-\mc{A}(\xi)$, and hence 
\begin{equation}\label{E_1}
\tilde{\alpha}_1(r,\xi)=\left(\frac{1}{r}-\frac{1}{r_h}\right)\mc{A}(\xi).
\end{equation}
This is exactly the result we obtained for a small electric field in Section \ref{SGF} \eqref{SGF_alpha}, and again this is exactly as expected, give that here $\alpha\sim O(\lambda)$. 
%
%
%

Turning to the magnetic field, we can integrate the equation for $\tilde{\omega}_1$ once, to give
\begin{equation}
\frac{\partial\tilde{\omega}_1}{\partial r}=\frac{\mc{L}(\xi)r-\mc{F}(\xi)}{r^3-r_h^3}
\end{equation}
for an arbitrary function $\mc{L}(\xi)$, and for this to be regular at $r=r_h$ we must take $\mc{L}(\xi)\equiv\frac{\mc{F}(\xi)}{r_h}$, so that
\begin{equation}
\frac{\partial\tilde{\omega}_1}{\partial r}=\frac{\mc{F}(\xi)}{r_h(r^2+rr_h+r_h^2)}.
\end{equation}
Integrating once more, we find
\begin{equation}\label{om1G}
\tilde{\omega}_1(r,\xi)=\mc{G}(\xi)+\frac{2\mc{F}(\xi)}{r_h^2\sqrt{3}}\tan^{-1}\left(\frac{2r+r_h}{r_h\sqrt{3}}\right),
\end{equation}
%
%
for an arbitrary function $\mc{G}$ which we can again specify by using \eqref{LLBCs}, obtaining
\begin{equation}\label{om1}
\tilde{\omega}_1(r,\xi)=\frac{\mc{F}(\xi)}{r_h^2\sqrt{3}}\left(2\tan^{-1}\left(\frac{2r+r_h}{r_h\sqrt{3}}\right)-\pi\right).
\end{equation}

In summary so far, we have found that to first order in $\lambda$, with fixed $|\Lambda|$ and $r_h$, the gauge fields are uniquely specified by the  two arbitrary functions $\omega_\infty(\xi)$ and $\mc{A}(\xi)$. However we must note that in order for those functions to correspond to genuine gauge field hair, they must agree with the form of the ans\"{a}tze \eqref{Ansatze}, meaning that they will have the forms
\begin{equation}\label{asymexp}
\mc{A}(\xi)=\slim_{j=1}^\infty\mc{A}_j\xi^{j}, \qquad \omega_\infty(\xi)=\sqrt{1-\xi^2}\slim_{j=0}^\infty\omega_{\infty,j}\xi^j,
\end{equation}
where $\{\mc{A}_j,\omega_{\infty,j}\}$ are two infinite sets of constants. 
Thus, Equations \eqref{E_1} and \eqref{om1} represent the appearance of genuine electric and magnetic gauge field hair in the Yang-Mills sector.

Now we consider the Einstein sector, which will show us what effect the hair has on the geometry, i.e. on the metric functions $\tilde{m}$ and $S$. We must examine terms of $O(\lambda^2)$ to see the influence of the magnetic field on the metric. We note that one would need to consider terms of $O(\lambda^3)$ and $O(\lambda^4)$ in the expansions for $\tilde{m}$ and $S$ (resp.) for the electric gauge field to manifest in the geometry itself. This is simply because we here require the gauge field to be small; continuing with this process of calculating the expansion terms yields expressions for $\tilde{m}_3$ and $\tilde{S}_4$ which depend partially on $\mc{A}(\xi)$.

The upshot is that the equations for the metric expansion functions at second order are similar to those in the purely magnetic case \cite{mavromatos_infinitely_2000}:
\begin{equation}\label{LL2eqn}
\begin{split}
\od{\hat{S}_2}{r}&=\frac{3\mc{C}_1H^2}{2rr_h^2},\\
\od{\hat{m}_2}{r}&=\frac{\mc{C}_1H}{4r_h}\left(\frac{1}{r}-\frac{1}{r_h}\right)+\frac{\mc{C}_2\sqrt{3}}{r^2r_h^2}\left(2\tan^{-1}\left(\frac{2r+r_h}{r_h\sqrt{3}}\right)-\pi\right),
\end{split}
\end{equation}
where we have defined
\begin{equation}
\begin{split}
H(r)&=\frac{1}{r^2+rr_h+r_h^2},\\
\mc{C}_1&=\int_{-1}^1\mc{F}^2d\xi,\\
\mc{C}_2&=\int_{-1}^1\left(\pd{}{\xi}\left(\omega_\infty^2+\xi^2\right)\right)\pd{}{\xi}\left(\omega_\infty\mc{F}\right)d\xi.\\
\end{split}
\end{equation}
The solutions to these equations are
\begin{equation}\label{LLorder2}
\begin{split}
\tilde{S}_2(r)&=\frac{\mc{C}_1}{4r_h^6}\left(3\ln( r^2H)+2r_h(r_h-r)H-\frac{10}{\sqrt{3}}\tan^{-1}\left(\frac{2r+r_h}{r_h\sqrt{3}}\right)\right)+\mc{K}_1,\\
\tilde{m}_2(r)&=\frac{1}{8r_h^3}\left(\mc{C}_1+12\mc{C}_2\right)\ln(r^2H)+\frac{\sqrt{3}\pi \mc{C}_2}{rr_h^2}\\
&-\frac{\sqrt{3}}{4r_h^3}\left(\mc{C}_1+4\mc{C}_2+\frac{8\mc{C}_2r_h}{r}\right)\tan^{-1}\left(\frac{2r+r_h}{r_h\sqrt{3}}\right)+\mc{K}_2,
\end{split}
\end{equation}
with arbitrary constants $\mc{K}_1$ and $\mc{K}_2$ which can be fixed using \eqref{LLBCs}. These solutions are globally regular.
%
%
%
%

In principle, there is no reason why this process could not be continued to arbitrary order in $\lambda$, where at the infinite limit we would obtain an exact solution to the field equations \eqref{EEs} and \eqref{YMs}. We wish to show that the above process of calculating the expansion functions can be continued indefinitely and that at each new order, the expansion functions will be determined entirely by $\Lambda$, $r_h$, and the two arbitrary functions $\omega_\infty(\xi)$, $\mc{A}(\xi)$ which appeared for the lowest order terms. 

Hence, we now substitute \eqref{laminfexp} into the field equations directly. Much algebraic manipulation yields the following recursive differential equations in the field variable expansion functions:
\begin{equation}\label{recur}
\begin{split}
\frac{d\tilde{m}_j}{dr}
&=\mc{X}_{1,j},\qquad \pd{}{r}\left(\left(\frac{r^3-r_h^3}{r}\right)\frac{\partial\tilde{\omega}_j}{\partial r}\right)=\mc{X}_{3,j}, \\
\frac{d\tilde{S}_j}{dr}
&=\mc{X}_{2,j},\qquad \pd{}{r}\left(r^2\frac{\partial\tilde{\alpha}_j}{\partial r}\right)=\mc{X}_{4,j},
\end{split}
\end{equation}
where $j\geq1$, 
%
%
and where the four functions $\mc{X}_{1,j}$ to $\mc{X}_{4,j}$ are extremely complicated expressions which 
for each integer $j$, depend only on the functions $\{\tilde{m}_k,\tilde{S}_k,\tilde{\omega}_k,\tilde{\alpha}_k\}$ for $k\leq j-1$.

Therefore, given that we know the solution at zeroeth order, we can in principle directly (recursively) solve \eqref{recur} to obtain the field variable expansions to any order, fixing arbitrary constants by requiring regularity at the event horizon, and also the boundary conditions \eqref{LLBCs}. This will ensure that the arbitrary functions or constants we get are \emph{at every order} expressed in terms of the two arbitrary functions $\omega_\infty(\xi)$ and $\mc{A}(\xi)$. Hence, the existence of exact solutions of the form \eqref{laminfexp}, which are completely defined by the two functions $\omega_\infty(\xi)$ and $\mc{A}(\xi)$, can be proven by straightforward mathematical induction. 

We may summarise the results of this Section in the following
\begin{thr}\label{lamlargethm}
	We regard the negative cosmological constant $\Lambda$ as very large and fixed and we regard the event horizon radius $r_h$ as constant. Then, to all orders of $\Lambda^{-1}$ in \eqref{laminfexp}, the functions $S$, $m$, $\omega$ and $\alpha$ are entirely specified by the two functions $\omega_\infty(\xi)$ and $\mc{A}(\xi)$, using the boundary conditions \eqref{LLBCs}, and the regularity of the event horizon. Thus, in principle, an exact solution may be obtained which is unique with respect to the asymptotic gauge degrees of freedom. 
\end{thr}

\section{Regular exterior black hole solutions}\label{reg}

Now that we have found some solutions to the field equations (\ref{EEs}, \ref{YMs}), it is important to prove that they all represent exterior fields of regular black holes, by calculating the associated curvature invariants. Relevant for the metric sector are the Riemann, Ricci and Kretschmann scalars, i.e. $R$, $R_{ab}R^{ab}$, and $R_{abcd}R^{abcd}$ respectively. In terms of our original metric \eqref{met}, these are given by
\begin{equation}\label{curvscal}
	\begin{split}
		R=&\,-\mu^{\prime\prime}-2\mu\frac{S^{\prime\prime}}{S}-3\mu^\prime\frac{S^\prime}{S}-\frac{4\mu^\prime}{r}-\frac{4\mu}{r}\frac{S^\prime}{S}+\frac{2(1-\mu)}{r^2},\\
		R_{ab}R^{ab}=&\,2\left(\frac{\mu^{\prime\prime}}{2}+\mu\frac{S^{\prime\prime}}{S}+\frac{3\mu^\prime S^\prime}{2S}+\frac{\mu^\prime}{r}\right)^{\! 2}+\frac{4\mu}{r}\frac{S^\prime}{S}\left(\frac{\mu^{\prime\prime}}{2}+\mu\frac{S^{\prime\prime}}{S}+\frac{3\mu^\prime S^\prime}{2S}+\frac{\mu^\prime}{r}\right)\\
		&+\frac{4\mu^2}{r^2}\left(\frac{S^\prime}{S}\right)^{\! 2}+\frac{2}{r^4}\left(r\mu^\prime+r\mu\frac{S^\prime}{S}-1+\mu\right)^{\! 2},\\
		R_{abcd}R^{abcd}=&\,\frac{4(1-\mu)^2}{r^4}+\frac{2\mu^{\prime 2}}{r^2}+4\left(\frac{\mu^{\prime\prime}}{2}+\mu\frac{S^{\prime\prime}}{S}+\frac{3\mu^\prime S^\prime}{2S}\right)^{\! 2}+\frac{8}{r^2}\left(\frac{\mu^\prime}{2}+\mu\frac{S^\prime}{S}\right)^{\! 2}.\\
	\end{split}
\end{equation}
In addition, we must calculate the Yang-Mills scalar curvature invariants $\mbox{Tr}\,F_{ab}F^{ab}$,  $\mbox{Tr}\,F_{ab}\tilde{F}^{ab}\in\R$, where $\tilde{F}^{ab}=\frac{1}{2}\varepsilon^{abcd}F_{cd}$ for $\varepsilon^{abcd}$ the Levi-Civita totally anti-symmetric symbol with $\varepsilon^{0123}=1$. This is to ensure the solution is of the black hole (i.e. non-extremal) type. We note that we are calculating the trace of the scalars $F_{ab}F^{ab}$, $F_{ab}\tilde{F}^{ab}\in\mk{su}(\infty)$, instead of just the scalars themselves -- the reason for this is that, noting the process in Section \ref{suinflimit} in which we take the limit $N\rar\infty$, the field strength tensor components acquire a factor of $N$ which is cancelled by inverse factors introduced by the trace (See \eqref{Traceop}), and therefore it is the trace that we expect to be well defined in this limit rather than the $\mk{su}(\infty)$-valued scalars themselves. In terms of 
the variables we defined in Section \ref{suinflimit}, 
these are given by:
\begin{equation}\label{YMcurvinv}
	\begin{split}
		\mbox{Tr}\,F_{ab}F^{ab}=&\,\frac{\eta}{2S^2}+\frac{\zeta}{2\mu S^2 r^2}-\frac{2\mu G}{r^2}-\frac{2P}{r^4},\\
		\mbox{Tr}\,F_{ab}\tilde{F}^{ab}=&\,\frac{\sin\theta}{2}\int\limits_{-1}^1\left(\pd{\alpha}{\xi}\pd{}{r}\left(\omega^2\right)-\pd{\alpha}{r}\pd{}{\xi}\left(\omega^2+\xi^2\right)\right)d\xi.
	\end{split}
\end{equation}
Finally, we recall that we also wish solutions to possess a non-extremal event horizon, with $\mu_h^\prime>0$, meaning that they should satisfy the criterion \eqref{mph}. 

We point out that upon investigating the solutions with small gauge fields in Section \ref{SGF}, we found that we must rely on results for purely magnetic $\essu$ solutions, and therefore we would need to take the limit $|\Lambda|\rar\infty$ since it is in that limit that characterising charges for purely magnetic $\essu$ solutions exist \cite{shepherd_characterizing_2012}. However, these are already covered by the solutions in Section \ref{LargeLambda} for gauge fields of general magnitude, so we ignore these solutions from here on in. Hence, we will take each of our solutions in turn: The non-trivial solutions with $|\Lambda|\rar\infty$ from Section \ref{LargeLambda}; and the trivial purely Abelian dyonic solutions from Section \ref{PureAb}.

\subsection{Solutions for large $|\Lambda|$}
	
First, let's consider the region where $r\geq r_h$ is a finite fixed value. We may examine $m$ and $S$ for the solution we derived in Section \ref{LargeLambda}, and see that these functions and their first derivatives are regular throughout $[r_h,\infty)$, since we assume regularity of the $\xi$-dependent functions $\omega_\infty$ and $\mc{H}$. Hence, we may calculate $m^{\prime\prime}$, $S^{\prime\prime}$, $\mu$, $\mu^\prime$ and $\mu^{\prime\prime}$, and see that these are also all regular for any finite $r\geq r_h$. In addition, $S$ is the only function in the denominators in \eqref{curvscal}, and for $|\Lambda|$ large enough, $S$ will be similar to 1 and hence non-zero everywhere. Therefore, it is clear that the quantities in \eqref{curvscal} are all finite if $r\geq r_h$ is finite -- the only singularity is at $r=0$. Then we can consider the asymptotic regime. Using the fact that as $r\rar\infty$, $S\sim 1$ and $\mu\sim-\frac{\Lambda r^2}{3}$, we can show that as $r\rar\infty$,
\begin{equation}\label{curvscal_r_inf}
	R=4\Lambda,\qquad R_{ab}R^{ab}=4\Lambda^2,\qquad R_{abcd}R^{abcd}=\frac{8\Lambda^2}{3},
\end{equation}
as expected for adS asymptotics. Therefore, noting that $|\Lambda|$ is large but still finite, the gravitational curvature scalars are also all regular as $r\rar\infty$. Hence they are regular throughout the range $[r_h,\infty)$.

Now we consider the Yang-Mills curvature invariants \eqref{YMcurvinv}. These depend on $\alpha$, $\omega$, and their first derivatives w.r.t. $r$ and $\xi$. Referring to Section \ref{LargeLambda}, for any finite $r\geq r_h$, these are all regular either by construction, or by assumption in the case of $\omega_\infty$ and $\mc{H}$. In addition, using asymptotic boundary conditions, we can 
%
%
conclude that $\mbox{Tr}\,F_{ab}F^{ab}\rar 0$ and $\mbox{Tr}\,F_{ab}\tilde{F}^{ab}\rar 0$ asymptotically. Therefore, the quantities in \eqref{YMcurvinv} are all globally regular. Since all relevant curvature scalars are globally regular, we can deduce that this solution represents a regular black hole exterior over the range $[r_h,\infty)$.

Lastly, we consider the regularity of the event horizon, given by the constraint \eqref{mph}. By applying boundary conditions at $r=r_h$ (Section \ref{BCrh}) to Equations \eqref{E_1} and \eqref{om1}, we can express the event horizon boundary data as
\begin{equation}\label{LLchargerh}
\omega_h(\xi)=\omega_\infty(\xi)-\frac{\pi\mc{F}(\xi)}{3\sqrt{3}\Lambda r_h^2}+O(\Lambda^{-2}),\qquad
\alpha^\prime_h(\xi)=-\frac{\mc{A}(\xi)}{\Lambda r_h^2}+ O(\Lambda^{-2});
\end{equation}
where again all the higher order terms are specified by $\omega_\infty(\xi)$ and $\mc{A}(\xi)$. Then, substituting those into \eqref{mph} gives the following constraint on solutions with regular event horizons:
\begin{equation}\label{regEH}
\frac{3}{8}\int\limits_{-1}^1\left(\pd{}{\xi}\left(\omega_\infty^2+\xi^2\right)\right)^2d\xi+O(\Lambda^{-1})<r_h^2(1-\Lambda r_h^2).
\end{equation}
%
The salient fact here is that (holding $r_h$ constant) if the magnitude of $\Lambda<0$ is large enough, the right hand side of \eqref{regEH} will be large and positive, and the terms of $O(\Lambda^{-1})$ on the left-hand side should be negligible. Therefore, it is possible to find some set of functions which could stand for $\omega_\infty(\xi)$ and which will satisfy the condition \eqref{regEH}. We will demonstrate this point in Section \ref{egsol}. Incidentally, the electric field only appears in the bound \eqref{regEH} at $O(\Lambda^{-2})$, and so has very little influence in the regularity of the event horizon. 

\subsection{Purely Abelian solutions}

The process here is similar and simpler. We recall that this solution is given in Section \ref{PureAb} by
\begin{equation}\label{PA}
	S=1,\qquad \omega=0,\qquad\alpha=\frac{\psi(\xi)}{r},\qquad \mu=1-\frac{2M}{r}-\frac{\Lambda r^2}{3}+\frac{\mc{Q}_E^2+1}{r^2}
\end{equation}
where $\mc{Q}_E^2$ is the electric charge, a constant defined in \eqref{QE2}, and $\psi(\xi)$ is an arbitrary function of $\xi$ which we assume to be regular in $[-1,1]$. We begin with the curvature invariants. We may substitute \eqref{PA} into \eqref{curvscal} to obtain lengthy expressions, and it can be easily observed that the curvature invariants are all regular for all $r\geq r_h$ finite. If we take the limit as $r\rar\infty$, we obtain the same results as we did for the previous case \eqref{curvscal_r_inf}, again as expected. 

The Yang-Mills curvature invariants \eqref{YMcurvinv} become
\begin{equation}
	\mbox{Tr}\,F_{ab}F^{ab}=\frac{2(\mc{Q}_E^2-1)}{3r^4},\qquad \mbox{Tr}\,F_{ab}\tilde{F}^{ab}=\frac{\sin\theta}{r^2}\int\limits_{-1}^1\xi\psi d\xi,
\end{equation}	
which are clearly regular for $r\geq r_h$ finite, and vanish as $r\rar\infty$.

As for the requirement of a regular event horizon, \eqref{mph} simplifies down to
\begin{equation}
	\mc{Q}_E^2+1<r_h^2(1-\Lambda r_h^2).
\end{equation}
%
So again, for fixed $r_h$, if $|\Lambda|$ is large enough and the charge $\mc{Q}_E^2$ is small enough, trivial purely Abelian solutions with a regular event horizon will exist. Hence, these solutions are also seen to be of the regular black hole type.

\section{An example solution family with $|\Lambda|$-large}\label{egsol}

\begin{center}
	\begin{figure}[h]
		\centering
		\includegraphics[scale=.6]{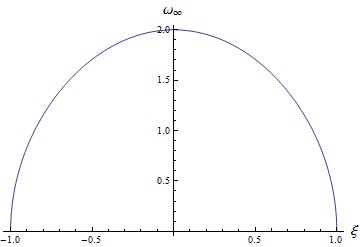}
		\caption{The example boundary data defined by \eqref{eg} with $W=2$.\label{A}}
	\end{figure}
\end{center}

\begin{center}
	\begin{figure}[h]
		\centering
		\includegraphics[scale=0.6]{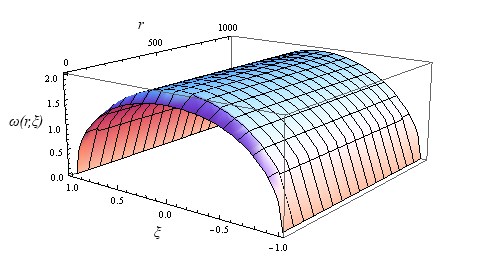}
		\caption{A plot of $\omega(r,\xi)$ for the example non-trivial $\suinf$ solution defined by \eqref{laminfexp} and \eqref{eg}, with $W=2$, $r_h=1$ and $\Lambda=-20$, including terms up to $O(\Lambda^{-1})$. 
		\label{C}}
	\end{figure}
\end{center}

\begin{center}
	\begin{figure}[h]
		\centering
		\includegraphics[scale=0.6]{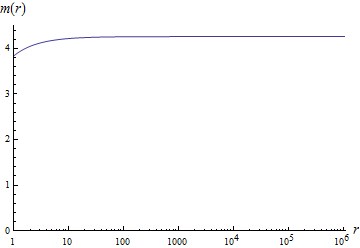}
		\caption{A plot of $m(r)$ for the example solution, including terms up to $O(\Lambda^{-2})$. 
			\label{m}}
	\end{figure}
\end{center}

\begin{center}
	\begin{figure}[h]
		\centering
		\includegraphics[scale=.55]{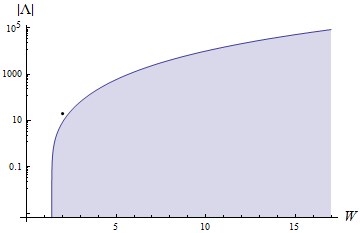}
		\caption{A parameter space plot of $|\Lambda|$ \emph{vs.} $W$ for the solutions defined by \eqref{eg}, with $r_h=1$. The area \emph{above} the shaded portion is the region satisfying \eqref{regEH}, where black holes with non-extremal event horizons may be found. The curve crosses the $W$-axis at $W=\sqrt{2}$. The marked point at $(2,20)$ is the location of the solution plotted in Figures \ref{C} and \ref{m}.}\label{B}
	\end{figure}
\end{center}

We present a simple concrete example of the solutions derived in Section \ref{LargeLambda}. We assume for a moment that the influence of the electric field is in fact negligible here, and for the magnetic gauge field, we let
\begin{equation}\label{eg}
\omega_\infty(\xi)=\frac{W}{\sqrt{2}}\sqrt{1-\xi^2}\sqrt{1+\frac{\cos\left(\frac{\pi\xi}{2}\right)}{1-\xi^2}}
\end{equation}
where $W>0$ is a constant which controls the `magnitude' of $\omega_\infty$, in that $W=\omega_\infty(0)$ is its maximum value. It can be seen that \eqref{eg} matches the form of the ansatz \eqref{asymexp}, and if it were expanded as such a power series, it would have an infinite series of expansion coefficients, which is necessary for the solution to be a genuine $\suinf$ solution. A plot of the function \eqref{eg} (with $W=2$) is given in Figure \ref{A}, and Figure \ref{C} is a plot of the global solution $\omega(r,\xi)$ which it generates, up to $O(\Lambda^{-1})$, and letting $r_h=1$ and $\Lambda=-20$. The mass function for the solution is plotted in Figure \ref{m} -- we can see that it is monotonic and has a finite limit at infinity, as we require. Finally, if we substitute \eqref{eg} into \eqref{regEH}, and again include terms up to $O(\Lambda^{-1})$, we may obtain a plot of $|\Lambda|$ against $W$ showing the region where solutions with non-extremal event horizons may be found (Figure \ref{B}) -- this region is the area \emph{above} the curve. The curve crosses the $W$-axis at $W=\sqrt{2}$, and the marked point is the location of the solution plotted in Figure \ref{C}. We notice that the set of values of $W$ which satisfy non-extremality \eqref{regEH} appears to grow without bound as $|\Lambda|$ grows, something we also saw in the case of $\sun$ \cite{baxter_existence_2008}. 

\section{Characterising solutions with a countable infinitude of global charges}\label{Charges}

In Sections \ref{SGF} and \ref{LargeLambda}, we discovered new classes of dyonic hairy black hole solutions which are specified by an infinite number of parameters. 
We now investigate these solutions to see whether we may also define \emph{characterising global charges} for these solutions, an issue which is raised by Bizon's modified ``No Hair'' Theorem. We will show here that one of the class of non-trivial solutions that we have derived, those in the regime where $|\Lambda|$ is large (Section \ref{LargeLambda}), can be entirely characterised by a set of charges, and furthermore, by no less than an infinite set of global charges.

By ``charges'', we are referring to a set of quantities defined asymptotically in terms of the gauge functions, which (together with $r_h$, $\Lambda$ and the ADM mass $M$) entirely characterise solutions of the field equations. In other words, once the charges are chosen, a unique solution to the field equations is entirely specified. In purely magnetic $\essu$ models, there are a couple of different possible definitions for the (single) Yang-Mills (YM) charge of the solution (e.g. \cite{corichi_mass_2000,mann_nonabelian_2006,kleihaus_rotating_2002}), and the existence of characterising YM charges for purely magnetic $\sun$ models for large $|\Lambda|$ and $r_h$ was established in \cite{shepherd_characterizing_2012}. We shall now derive expressions for the charges in the case of $\suinf$, and relate our expressions to the trivial and the non-trivial solutions we have described in this work, as well as charges found for $\sun$ solutions \cite{shepherd_characterizing_2012}. It should be pointed out that the following method is not the only possible way to define charges, see for instance \cite{corichi_mass_2000}, but it is one that gives us results in agreement with embedded solutions.

Here we follow \cite{chrusciel_global_1987,shepherd_characterizing_2012}. We let $X$ be an element of the Cartan subalgebra $\mk{h}$. Then, corresponding to each $X\in\mk{h}$ we may define scalar charges as follows:
\begin{equation}\label{QX}
\mc{Q}(X)=\frac{1}{4\pi}\sup_{g(r)\in\mk{h}}\,K\left(X,\int_{S_\infty}g^{-1}Fg\right),
\end{equation}
where the integral is taken over a sphere at infinity, and $K(\cdot\, ,\cdot)$ is the usual Killing form on the Lie algebra. The quantity $F$ is related to the appropriate component of the gauge field strength tensor $F_{\mu\nu}$ for each gauge field sector: for the magnetic charge, we define $F_M\equiv F_{\theta\phi}$; and for the electric charge, we define $F_E\equiv F_{tr}\sin\theta\propto\star F_{\theta\phi}$, the Hodge dual of $F_{\theta\phi}$ \cite{corichi_mass_2000}. Using the correspondence given in Section \ref{AnsFE}, these become
\begin{equation}\label{F}
F_E=-\frac{iN}{2}\pd{\alpha}{r}\sin\theta,\quad F_M=-\frac{iN}{2}\pd{}{\xi}\left(\omega^2+\xi^2\right)\sin\theta,
\end{equation}
(where we note that the factors of $N$ end up being cancelled out by the trace \eqref{Traceop}). The supremum in \eqref{QX} is taken over all elements $g(r)\in\mk{h}$ such that 
%
$g(r)=\exp[f(r)\sigma]$, 
%
where $f(r)$ is a scalar function of $r$ and $\sigma$ is a constant Lie algebra element. 
It can be shown in general \cite{chrusciel_global_1987} that the integrand $g^{-1}Fg$ in \eqref{QX} takes its maximum value when $g^{-1}Fg\in\mk{h}$. However, it may be noticed by writing them out that $F_{tr}$ and $F_{\theta\phi}$ are both diagonal matrices in $\sun$, and hence are both elements of $\mk{h}$, meaning that we are free to choose $g(r)=e$ for $e$ the identity element in $\sun$. Hence, in the $\sun$ case,  
we may use \eqref{QX} to define
\begin{equation}\label{QX2}
\mc{Q}_E(X)\equiv\lim_{r\rar\infty}K\left(X,F_{tr}\right),\,\, X\in\mk{h}
\end{equation}
(with an analogous definition for $Q_M(X)$), and we choose $Q_E(X_E)$ to be the electric charge of the solution, and $Q_M(X_M)$ to be the magnetic charge. If we were working with a finite gauge group we should now wish to find an appropriate basis for $X_E$ and $X_M$ in the Cartan subalgebra. However, for the case of $\suinf$, we expect for each charge to be defined by a single continuous function. 
Therefore, we just need to translate \eqref{QX2} into an expression for $\suinf$ black holes, and find two functions of $\xi$ to serve as the elements $X_E$ and $X_M$.

For $\sun$, the Killing form $K(X,Y)=\mbox{Tr}(\mbox{ad}\,X,\mbox{ad}\,Y)$ is defined in terms of the adjoint map on Lie algebra elements $\mbox{ad}\,X\equiv[X,\,\cdot\,\,]$ -- i.e. in the adjoint representation, it is just the matrix trace of the product, $\mbox{Tr}(\tilde{X}\tilde{Y})$ for $\tilde{X},\tilde{Y}$ the matrix representations of the $X,Y\in\sun$. Noting then the process of taking the limit $N\rar\infty$ from Section \ref{suinflimit}, we define
%
%
%
\begin{equation}\label{chargedefs}
\begin{split}
\mc{Q}^2_E\equiv&\mc{Q}_E(X_E)\equiv\frac{1}{2N}\lim_{r\rar\infty}\int\limits_{-1}^1-\frac{i}{2}X_E\pd{\alpha}{r}d\xi,\\ \mc{Q}^2_M\equiv& \mc{Q}_M(X_M)\equiv\frac{1}{2N}\lim_{r\rar\infty}\int\limits_{-1}^1-\frac{i}{2}X_M\left(\pd{}{\xi}\left(\omega^2+\xi^2\right)\right)d\xi.
\end{split}
\end{equation}
We define the square of the charge, rather than the charge itself as in \cite{chrusciel_global_1987,shepherd_characterizing_2012}, since the notation will then conveniently accord with notation used for simpler cases where the total charge is always expressed as a square (e.g. the Reissner-N\"{o}rdstrom case, and see also \cite{baxter_existence_2018,shepherd_characterizing_2012}). 

The choice of our elements $X_E,X_M\in\mk{h}$ will be restricted by requiring that the squared charges reduce correctly to those of the trivial solutions in the correct regimes when we use units in which $\kappa=3$ (Section \ref{Embed}), and that we also obtain the correct total squared charge for purely magnetic $\sun$ solutions \cite{shepherd_characterizing_2012} if we use the method of lines gives in Section \ref{Embedsun} and use units in which $\kappa=2$. We remind the reader that the value of $\kappa$ depends simply upon a convenient choice of units. Furthermore, we wish our charges to coincide with a total \emph{effective charge} $Q_{eff}^2$, defined as the charge that plays the same role in the geometry as the usual Abelian charge (Section \ref{Embed}). This means that asymptotically, we require the mass function to be given by
\begin{equation}\label{Qeff}
m(r)=M-\frac{Q_{eff}^2}{2r}+O(r^{-2}).
\end{equation}
We will return to this point later.

To meet these requirements, we choose 
%
$X_E=-\kappa r^4 F_{tr}$ and $X_M=-\kappa F_{\theta\phi}(\sin\theta)^{-1}$ 
%
so that it is clear that $X_E,X_M\in\mk{h}$, and obtain
\begin{equation}\label{charges}
\mc{Q}^2_E=\frac{1}{2}\int\limits_{-1}^1q_E^2 d\xi,\qquad
\mc{Q}^2_M=\frac{1}{2}\int\limits_{-1}^1q_M^2d\xi,
\end{equation}
where
\begin{equation}\label{chargebij}
q_E(\xi)\equiv\frac{\sqrt{\kappa}}{2}\mc{A}(\xi),\qquad q_M(\xi)\equiv\frac{\sqrt{\kappa}}{2}\od{}{\xi}\left(\omega^2_\infty(\xi)+\xi^2\right),
\end{equation}
and we recall that $\omega_\infty(\xi)$ and $\mc{A}(\xi)$ are defined in Section \ref{BCinf}. We also note that we are able to ignore a trivial sign ambiguity in $\omega_\infty$ due to the symmetry of the field equations $\omega\mapsto-\omega$ (see Section \ref{symm}). Thus, only the asymptotic data functions $\omega_\infty(\xi)$ and $\mc{A}(\xi)$ are required to specify the charge functions $q_E(\xi)$ and $q_M(\xi)$, and hence the total charges.

We point out that we have made a distinction here between the \emph{charge functions} $q_E(\xi)$ and $q_M(\xi)$, and the \emph{total charges}, the constants $\mc{Q}_E^2$ and $\mc{Q}_M^2$, which are the integrals of their squares. This in perfect analogy with the total magnetic charge for the $\sun$ solution $Q^2$ being expressed as a sum over $N-1$ squared charges $Q_j^2$, one for each gauge degree of freedom \cite{shepherd_characterizing_2012}. 
It is therefore the charge functions $q_E(\xi)$ and $q_M(\xi)$ which we expect to characterise the solutions. Also, it is clear from \eqref{F} and \eqref{chargebij} that $q_E,q_M\in\mk{h}$, so we also have the constraints 
\begin{equation}\label{intq}
\int_{-1}^1q_E d\xi=\int_{-1}^1q_M d\xi=0.
\end{equation}
%
We may invert the expression for $q_M$ in \eqref{chargebij} by integrating, rearranging, and using the boundary condition $\omega(r,1)=0$ to give
\begin{equation}\label{invert}
\omega_\infty^2(\xi)=1-\xi^2-\frac{2}{\sqrt{\kappa}}\int\limits_\xi^1 q_M(z)dz.
\end{equation}
Note that as in the $\essu$ case \cite{mann_nonabelian_2006} we are allowing for a trivial sign ambiguity in $\omega$ (See Equation \eqref{mirsym}). Also, it can be noticed that we could have chosen to use the other boundary condition $\omega(r,-1)=0$ and obtained
\begin{equation}\label{invert2}
\omega_\infty^2(\xi)=1-\xi^2+\frac{2}{\sqrt{\kappa}}\int\limits_{-1}^\xi q_M(z)dz,
\end{equation}
but with \eqref{intq} we can see that $\int_{-1}^\xi q_M(z)dz=-\int_{\xi}^1 q_M(z)dz$, and so \eqref{invert} and \eqref{invert2} give the same result for $\omega_\infty^2$. Therefore, Equations \eqref{chargebij} clearly define bijections between the charge functions $\{q_E,q_M\}$ and the asymptotic data $\{\mc{A},\omega_\infty\}$. It can also be noted that the charge functions themselves are specified by a countably infinite number of parameters, which we could take as coming from the expansions \eqref{asymexp} of the gauge fields; alternatively, noting the forms of the polynomials we will get by substituting \eqref{asymexp} into \eqref{chargebij}, we can write $q_E(\xi)$ and $q_M(\xi)$ for $\xi\in\mc{I}$ as
\begin{equation}\label{qexp}
q_E(\xi)=\slim_{j=0}^\infty q_{E,j}\xi^j,\qquad q_M(\xi)=\slim_{j=0}^\infty q_{M,j}\xi^{j}.
\end{equation}
Then, the individual charges would be the infinite set of expansion coefficients $\{q_{E,j},q_{M,j}\}$. Finally, it can be seen that if $q_E(\xi)=q_M(\xi)\equiv0$, i.e. the solution has zero charge, then \eqref{invert} recovers the asymptotic data of the SAdS solution (Section \ref{EmbedSAdS}).

The expressions in \eqref{charges} and \eqref{chargebij} reduce correctly to expressions for the charges of trivial solutions in Section \ref{Embed}, and the squared magnetic charge $\mc{Q}_M^2$ \eqref{charges} matches that in the case of purely magnetic $\sun$ solutions \cite{shepherd_characterizing_2012}, using the method of lines in Section \ref{Embedsun}. Note that $\kappa$ is included in the definitions \eqref{chargebij} to allow for the different cases, but note that we could also just keep using the units where $\kappa=3$, in which case the defined charges would be at most proportional to those in the $\sun$ purely magnetic case, and this is still acceptable from a uniqueness perspective.

The main point here is that if the solutions that we have found can be shown to be entirely specified by their asymptotic boundary functions $\{\mc{A},\omega_\infty\}$, then given that the maps \eqref{charges} from the asymptotic functions $\{\mc{A},\omega_\infty\}$ to the charge functions are bijections, the solutions we found would be entirely characterised by their charge functions $q_E$ and $q_M$, as required by Bizon's modified ``No Hair" theorem. However, due to the ans\"{a}tze \eqref{Ansatze}, each non-trivial gauge field function \emph{must} be expressed in terms of a \emph{countably infinite} number of coefficients in general. 
As we shall explain in the conclusion, this could be a rather significant result for the status of Bizon's modified ``No Hair'' conjecture. 

We  may therefore prove the following, omitting the solutions with small gauge fields (Section \ref{SGF}), since we discovered that solutions for these only exist in the regime $|\Lambda|\rar\infty$:
\begin{thr}\label{chargethm}
	We fix the values of $r_h$ and $\Lambda$. Then, global charge functions $q_M(\xi)$ and $q_E(\xi)$ may be found which entirely and uniquely characterise the solutions to the field equations \eqref{EEs}, \eqref{YMs} explicitly in terms of the asymptotic boundary functions $\omega_\infty(\xi)$ and $\mc{A}(\xi)$, in the following cases: the trivial purely Abelian case (Section \ref{PureAb}); 
	and in the regime where $\Lambda<0$, $|\Lambda|\rar\infty$ (Section \ref{LargeLambda}), including at least some solutions which respect the criterion for a regular event horizon \eqref{regEH}. In turn, the charge functions are each uniquely determined by a countably infinite set of parameters.
\end{thr}

\textbf{Proof} We have defined maps \eqref{chargebij} in which the charge functions uniquely specify the asymptotic functions $\{\mc{A}(\xi),\omega_\infty(\xi)\}$. What we show now is that knowledge of the functions $\{\mc{A}(\xi),\omega_\infty(\xi)\}$ entirely specifies the solution to which that asymptotic data belongs. It should be noted that since $S\rar1$ in the limit $r\rar\infty$ \eqref{infexp}, we can concentrate on the other three functions $m$, $\omega$ and $\alpha$. Also, below we take $\kappa=3$, so that the charges reduce correctly to the charge for purely magnetic $\essu$ solutions. We point out that analogous results for $\essu$ dyonic field equations do not currently exist, though the existence of defining charges in that case is heavily implied by the following results, due to the existence of the $\essu$ embedding (Section \ref{Embedsu2}).

\emph{Trivial purely Abelian solutions (Section \ref{PureAb}):} Here, we fix $\omega\equiv0$, which implies $\omega_\infty=\mc{W}\equiv0$; and comparing \eqref{infexp} with \eqref{pureabsol}, we let $\alpha_\infty\equiv0$, $\mc{A}(\xi)=\psi(\xi)$. The gauge sector of the solution is entirely determined by $\psi(\xi)$. The charge functions are given by $q_E(\xi)=\frac{\sqrt{3}}{2}\psi(\xi)$, which makes perfect sense if we recall that \eqref{pureabsol} defines a Coloumb-like potential; and $q_M(\xi)=\sqrt{3}\xi$, meaning that $\mc{Q}^2_M=1$. The ADM mass $M$ can be determined using \eqref{PureAbM} and boundary conditions at the event horizon as $M=m_h+(\mc{Q}_E^2+1)/(2r_h)$, with $Q^2_E$ defined in Equation \eqref{QE2}. Therefore, the charges uniquely determine the asymptotic data which then determine the solution. Finally, it can be seen from the form of $m(r)$ \eqref{PureAbM} that our total charge $\mc{Q}^2$ coincides with our defined effective charge $Q^2_{eff}$ \eqref{Qeff}.

\emph{Solutions for $|\Lambda|$ large (Section \ref{LargeLambda}):} First of all, we can see that our charge definitions coincide in this regime with the total effective charge defined by \eqref{Qeff}, where we let $Q_{eff}^2=Q^2_{E,eff}+Q^2_{M,eff}$ in which $Q^2_{E,eff}$ and $Q^2_{M,eff}$ are the effective electric and magnetic charges respectively. Examining \eqref{infexp} and \eqref{m1}, we see that $Q_{eff}^2=-2m_1$; and it can be seen (below) with \eqref{LLchargeinfbh} that $\mc{W}^2\sim O(\Lambda^{-2})$. 
Therefore, considering $m_1$ \eqref{m1} in this limit i.e. considering all terms of $O(\Lambda^{-1})$ and smaller to be negligible, we can consistently define
\begin{equation}
Q^2_{E,eff}=\frac{3}{8}\int^1_{-1}\mc{A}^2d\xi,\qquad Q^2_{M,eff}=\frac{3}{8}\int^1_{-1}\left(\od{}{\xi}\left(\omega_\infty^2+\xi^2\right)\right)^2d\xi.
\end{equation}
Thus, in the limit of large-$|\Lambda|$, the expression for the effective charge coincides with our charge definitions given in Equations \eqref{charges} and \eqref{chargebij}.

Using Theorem \ref{lamlargethm}, the solutions we constructed in Section \ref{LargeLambda} entirely determined upon the choice of the arbitrary functions $\omega_\infty$ and $\mc{A}$; that is, if we specify functions $\{\mc{A},\omega_\infty\}$, this entirely specifies a solution. Thence, given that the charge functions uniquely specify the asymptotic data, the charge functions $\{q_E,q_M\}$ specify the unique solution to which they belong.  

We can make this absolutely clear by showing that the asymptotic boundary functions $\{\mc{A},\omega_\infty\}$ (along with $\Lambda$ and $r_h$) entirely define all the rest of the boundary data, both as $r\rar\infty$ \emph{and} at $r=r_h$, and hence $\{\mc{A},\omega_\infty\}$ unquestionably fixes the entire solution. We already have expressions for the event horizon boundary data \eqref{LLchargerh}, 
and the other parameters which determine the gauge sector asymptotically, i.e. the ADM mass $M$ and the functions $\{\mc{W},\alpha_\infty\}$ (See Sections \ref{BCinf} and \ref{LargeLambda}), can be written as follows:
\begin{equation}\label{LLchargeinfbh}
\begin{split}
\alpha_\infty(\xi)&=
-\frac{\mc{A}(\xi)}{\Lambda r_h}+O(\Lambda^{-2}),\\
\mc{W}(\xi)&=
-\frac{\mc{F}(\xi)}{\Lambda r_h}+O(\Lambda^{-2}),\\
M&=-\frac{\Lambda r_h^3}{6}+\frac{r_h}{2}+\frac{K_1}{2r_h}+O(\Lambda^{-1}).
\end{split}
\end{equation}
In Equations \eqref{LLchargerh} and \eqref{LLchargeinfbh}, $\mc{F}(\xi)$ and $K_1$ are given in \eqref{F1andK1} and are fixed by $\omega_\infty(\xi)$ -- in fact, it can be seen that 
\begin{equation}
K_1=\mc{Q}^2_M,\qquad\mc{F}=\sqrt{3}\omega_\infty\od{q_M}{\xi}.
\end{equation}
The order $O(\Lambda^{-1})$ term in $M$ \eqref{LLchargeinfbh} can be calculated by taking the asymptotic limit of \eqref{LLorder2}, and is expressed in terms of $\mc{F}(\xi)$ and hence $\omega_\infty(\xi)$. In addition, by Theorem \ref{lamlargethm}, all terms of $O(\Lambda^{-2})$ and lower in the Equations \eqref{recur}, and hence in Equations \eqref{LLchargerh} and \eqref{LLchargeinfbh}, can also in principle be calculated from the field variable expansion terms of $O(\Lambda^{-1})$ and $O(1)$, and the constants and functions of integration may be fixed by the boundary conditions at $r=r_h$ from Section \ref{BCrh}. Hence, all higher order terms above depend solely on the functions $\{\mc{A},\omega_\infty\}$. Therefore, noting \eqref{chargebij} and \eqref{invert}, this solution is entirely fixed by the values of $r_h$, $\Lambda$, and the functions $q_E(\xi)$ and $q_M(\xi)$.

Finally, given \eqref{asymexp}, we can see for both of the above cases that the functions $\mc{A}(\xi)$ and $\omega_\infty(\xi)$ are uniquely determined by a countably infinite set of parameters, i.e. the constant coefficients $\{\mc{A}_j,\omega_{\infty,j}\}$ which appear when the functions $\mc{A}$ and $\omega_{\infty}$ are expressed in the form \eqref{asymexp}. Hence, so are the charge functions $q_E$ and $q_M$ \eqref{chargebij}. $\Box$

\section{Conclusions}\label{Conc}

In this work, we have derived field equations for asymptotically AdS, dyonic $\suinf$ EYM models, and presented two classes of analytical black hole solutions to these equations: i) Solutions where the gauge fields are small, but for a general value of $\Lambda<0$; and ii) solutions for a large value of $|\Lambda|$, where the gauge fields are of general magnitude. In addition, we defined expressions for the charges of $\suinf$ EYM solutions. We also found the pleasing result that the solutions from class ii), where $|\Lambda|\rar\infty$, can be characterised uniquely by asymptotically measured charge functions. Due to Theorem \ref{chargethm}, the charge functions are each uniquely characterised by an \emph{infinite} number of parameters. Bizon's modified ``No hair'' conjecture states that a black hole in a given matter model is defined by a \emph{finite} number of global charges. This means we may have found solutions that fall outside of its scope. Noting closely the language, the Theorem speaks of ``stable black holes'', so a linear stability analysis of time-dependent $\suinf$ EYM field equations system is a crucial next step, and the work for this is already in progress \cite{baxter_inprep}. To update the language of such an important conjecture would indeed be exciting.

The models presented here are classical field configurations. In order to investigate the issue of quantum decoherence during Hawking evaporation and hence shed light on black hole information loss, it would be necessary to quantise the system, taking spacetime fluctuations into account. A method was devised in \cite{mavromatos_aspects_1996} for quantifying the entropy production (i.e. loss of information) of an evaporating $\essu$ EYM black hole minimally-coupled to a scalar field, but there are reasons to believe that the holographic properties of the system may not survive such a process if the scalar field is minimally coupled. Indeed, the holographic properties pertain to the solutions with small gauge fields, i.e. those nearby RNAdS solutions, and it is known that entropy production for RNAdS black holes is non-trivial \cite{cai__1996}. However, it is perfectly possible that if we include a scalar field which is non-minimally coupled to the (holographic) gauge degrees of freedom, the situation may be completely different. This is a problem to which we may return.

A further interesting area of potential research would be to extend this work to so-called `topological solutions' -- 4D solutions in surface-symmetric spacetimes, i.e. which are foliated by 2D surfaces of constant Gaussian curvature. This is already a subject of ongoing interest \cite{shepherd_black_2017, baxter_topological_2016}. A knowledge of topological $\suinf$ dyons would potentially be very useful to holographic superconductors, since in the limit $N\rar\infty$ the dual CFT which corresponds to it will be `exact' in the sense described in Section \ref{Intro}. However, these models are tricky to construct, in that most stages of the construction given in Section \ref{AnsFE} will not apply to non-spherically symmetric models, so this is likely to be quite challenging.

\appendix
%
%

\newpage
\section*{References}
%
%
%

\begin{thebibliography}{10}
	
	\bibitem{israel_300_1987}
	W.~Israel and S.~Hawking.
	\newblock {\em 300 years of {Gravitation}}.
	\newblock Cambridge Uni Press, 1987.
	
	\bibitem{israel_event_1968}
	W.~Israel.
	\newblock {\em Comm. Math. Phys.}, 8(3):245--260, 1968.
	
	\bibitem{israel_event_1967}
	W.~Israel.
	\newblock {\em Phys. Rev.}, 164(5):1776--1779, 1967.
	
	\bibitem{bizon_gravitating_1994}
	P.~Bizon.
	\newblock {\em Acta Phys. Polon. B}, 25:877--98, 1994.
	
	\bibitem{bartnik_particle-like_1988}
	R.~Bartnik and J.~McKinnon.
	\newblock {\em Phys. Rev. Lett.}, 61:141--144, 1988.
	
	\bibitem{volkov_number_1995}
	M.~S. Volkov, O.~Brodbeck, G.~Lavrelashvili, and N.~Straumann.
	\newblock {\em Phys. Lett. B}, 349:438--42, 1995.
	
	\bibitem{zhou_instability_1990}
	Z.~H. Zhou.
	\newblock {\em Phys. Lett. B}, 237:353, 1990.
	
	\bibitem{lavrelashvili_remark_1995}
	G.~Lavrelashvili and D.~Maison.
	\newblock {\em Phys. Lett. B}, 343(1-4):214--217, 1995.
	
	\bibitem{brodbeck_instability_1996}
	O.~Brodbeck and N.~Straumann.
	\newblock {\em Jour. Math. Phys.}, 37(3):1414--1433, 1996.
	
	\bibitem{winstanley_existence_1999}
	E.~Winstanley.
	\newblock {\em Class. Quant. Grav.}, 16(6):1963--1978, 1999.
	
	\bibitem{sarbach_linear_2001}
	O.~Sarbach and E.~Winstanley.
	\newblock {\em Class. Quant. Grav.}, 18:2125--2146, 2001.
	
	\bibitem{winstanley_linear_2002}
	E.~Winstanley and O.~Sarbach.
	\newblock {\em Class. Quant. Grav.}, 19:689--724, 2002.
	
	\bibitem{radu_static_2004}
	E.~Radu and E.~Winstanley.
	\newblock {\em Phys. Rev. D}, 70(8):084023, 2004.
	
	\bibitem{baxter_topological_2016}
	J.~E. Baxter and E.~Winstanley.
	\newblock {\em Phys. Lett. B}, 753:268--273, 2016.
	
	\bibitem{winstanley_instability_2016}
	E.~Winstanley.
	\newblock {\em Phys. Lett. B}, 758:239--243, 2016.
	
	\bibitem{ponglertsakul_solitons_2016}
	S.~Ponglertsakul and E.~Winstanley.
	\newblock {\em Phys. Rev. D}, 94(4):044048, 2016.
	
	\bibitem{baxter_stability_2016}
	J.~E. Baxter and E.~Winstanley.
	\newblock {\em Jour. Math. Phys.}, 57(2):022506, 2016.
	
	\bibitem{baxter_abundant_2008}
	J.~E. Baxter, M.~Helbling, and E.~Winstanley.
	\newblock {\em Phys. Rev. Lett.}, 100(1):011301, 2008.
	
	\bibitem{baxter_existence_2016}
	J.~E. Baxter.
	\newblock {\em Jour. Math. Phys.}, 57(2):022505, 2016.
	
	\bibitem{baxter_existence_2018}
	J.~E. Baxter.
	\newblock {\em Jour. Math. Phys.}, 59(5):052502, 2018.
	
	\bibitem{nolan_existence_2012}
	B.~C. Nolan and E.~Winstanley.
	\newblock {\em Class. Quant. Grav.}, 29(23):235024, 2012.
	
	\bibitem{winstanley_menagerie_2015}
	E.~Winstanley.
	\newblock {\em Proceedings of the Second Karl Schwarzschild
		Meeting, Frankfurt, 20-24 July 2015}, Springer Proceedings in Physics, 2015.
	
	\bibitem{nolan_stability_2016}
	B.~C. Nolan and E.~Winstanley.
	\newblock {\em Class. Quant. Grav.}, 33(4):045003, 2016.
	
	\bibitem{maldacena_large_1998}
	J.~M. Maldacena.
	\newblock {\em Adv. Theor. Math. Phys.}, 2:231--52, 1998.
	
	\bibitem{hendi_holographical_2018}
	S.~H. Hendi, N.~Riazi, and S.~Panahiyan.
	\newblock {\em Ann. der Phys.}, 530(2):1700211, 2018.
	
	\bibitem{cai_introduction_2015}
	R-G. Cai, L.~Li, L-F Li, and R-Q. Yang.
	\newblock {\em Sci. Chi. Phys., Mech. \& Astr.}, 58(6):060401, 2015.
	
	\bibitem{shepherd_black_2017}
	B.~L. Shepherd and E.~Winstanley.
	\newblock {\em Jour. High Energy Physics}, 2017(1), 1701 065, 2017.
	
	\bibitem{bakas_large_1989}
	I.~Bakas.
	\newblock {\em Phys. Lett.}, B228:57, 1989.
	
	\bibitem{sezgin_area-preserving_1992}
	E.~Sezgin.
	\newblock {\em 1991 summer school in high energy physics and cosmology. V. 2},
	1992.
	
	\bibitem{pope_higher-spin_1990}
	C.~N. Pope and X.~Shen.
	\newblock {\em Phys. Lett. B}, 236(1):21--26, 1990.
	
%
%
	\bibitem{ellis_w_infty_2016}
	J.~Ellis, N.~E. Mavromatos, and D. V. Nanopoulos.
	\newblock {\em Phys. Rev. D}, 94(2):025007, 2016.
	
	\bibitem{antoniadis_proceedings_2015}
	I.~Antoniadis, G.~K. Leontaris, and K.~Tamvakis.
	\newblock {\em PoS}, PLANCK 2015, 2015.
	
	\bibitem{hawking_soft_2016}
	S.~W. Hawking, M.~J. Perry, and A.~Strominger.
	\newblock {\em Phys. Rev. Lett.}, 116:231301, 2016.
	
	\bibitem{haco_black_2018}
	S.~Haco, S.~W. Hawking, M.~J. Perry, and A.~Strominger.
	\newblock {\em J. High Energy Phys.}, 2018 098, 2018.
	
	\bibitem{mavromatos_infinitely_2000}
	N.~E. Mavromatos and E.~Winstanley.
	\newblock {\em Class. Quant. Grav.}, 17(7):1595--1611, 2000.
	
	\bibitem{hoppe_quantum_1982}
	J.~Hoppe.
	\newblock Ph{D} thesis, Massachusetts Institute of Technology, 1982.
	
	\bibitem{floratos_note_1989}
	E.~G. Floratos, J.~Iliopoulos, and G.~Tiktopoulos.
	\newblock {\em Phys. Lett. B}, 217(3):285--288, 1989.
	
	\bibitem{baxter_soliton_2007}
	J.~E. Baxter, M.~Helbling, and E.~Winstanley.
	\newblock {\em Phys. Rev. D}, 76(10):104017, 2007.
	
	\bibitem{shepherd_characterizing_2012}
	B.~L. Shepherd and E.~Winstanley.
	\newblock {\em Class. Quant. Grav.}, 29(15):155004, 2012.
	
	\bibitem{baxter_existence_2008}
	J.~E. Baxter and E.~Winstanley.
	\newblock {\em Class. Quant. Grav.}, 25(24):245014, 2008.
	
	\bibitem{graham_einstein_1991}
	C.~R. Graham and J.~M. Lee.
	\newblock {\em Adv. in Math.}, 87(2):186--225, 1991.
	
	\bibitem{hawking_thermodynamics_1983}
	S.~W. Hawking and D.~N. Page.
	\newblock {\em Comm. Math. Phys.}, 87:577, 1983.
	
	\bibitem{witten_anti-sitter_1998}
	E.~Witten.
	\newblock {\em Adv. Theor. Math. Phys.}, 2:253--291, 1998.
	
	\bibitem{corichi_mass_2000}
	A.~Corichi and D.~Sudarsky.
	\newblock {\em Phys. Rev. D}, 61(10):101501, 2000.
	
	\bibitem{mann_nonabelian_2006}
	R.~B. Mann, E.~Radu, and D.~H. Tchrakian.
	\newblock {\em Phys. Rev. D}, 74(6):064015, 2006.
	
	\bibitem{kleihaus_rotating_2002}
	B.~Kleihaus, J.~Kunz, and F.~Navarro-Lerida.
	\newblock {\em Phys. Rev. D}, 66(10):104001, 2002.
	
	\bibitem{chrusciel_global_1987}
	P.~T. Chrusciel and W.~Kondracki.
	\newblock {\em Phys. Rev. D}, 36(6):1874--1881, 1987.
	
	\bibitem{baxter_inprep}
	J.~E. Baxter.
	\newblock {\em Work in preparation.}
	
	\bibitem{mavromatos_aspects_1996}
	N.~E. Mavromatos and E.~Winstanley.
	\newblock {\em Phys. Rev. D}, 53(6):3190--3214, 1996.
	
	\bibitem{cai__1996}
	R-G. Cai and Y-Z. Zhang.
	\newblock {\em Mod. Phys. Lett. A}, 11:2027, 1996.
	
\end{thebibliography}

\bibliographystyle{unsrt}

\end{document}